\documentclass[twocolumn]{aastex6}

\pdfoutput=1

\usepackage{mathtools}
\usepackage[normalem]{ulem}
\usepackage{xcolor}
\usepackage{afterpage}

\newcommand{\cholla}{\textit{Cholla}~}
\newcommand{\chollans}{\textit{Cholla}}
\newcommand{\tcc}{$t_\mathrm{cc}$~}
\newcommand{\tccns}{$t_\mathrm{cc}$}
\newcommand{\turb}{turbulent }

\begin{document}

\title{Hydrodynamical Coupling of Mass and Momentum in Multiphase Galactic Winds}

\author{Evan E. Schneider\altaffilmark{1} and Brant E. Robertson\altaffilmark{2}}
\altaffiltext{1}{Steward Observatory, University of Arizona, 933 North Cherry Avenue, Tucson, AZ 85721, USA}
\altaffiltext{2}{Department of Astronomy and Astrophysics, University of California, Santa Cruz, 1156 High Street, Santa Cruz, CA 95064 USA}

\begin{abstract}
Using a set of high resolution hydrodynamical simulations run with the \cholla code, we investigate how mass and momentum couple to the multiphase components of galactic winds. The simulations model the interaction between
a hot wind driven by supernova explosions and a cooler, denser cloud of interstellar or circumgalactic media. By resolving scales of $\Delta x<0.1$ pc over $>100$ pc distances our calculations capture how the cloud disruption leads to a distribution of densities and temperatures in the resulting multiphase outflow, and quantify the mass and momentum associated with each phase. We find the multiphase wind contains comparable mass and momenta in phases over a wide range of densities and temperatures extending from the hot wind ($n \approx 10^{-2.5}$~$\mathrm{cm}^{-3}$, $T \approx 10^{6.5}$~K) to the coldest components ($n \approx 10^2$ $\mathrm{cm}^{-3}$, $T \approx 10^2$ K). We further find that the momentum distributes roughly in proportion to the mass in each phase, and the mass-loading of the hot phase by the destruction of cold, dense material is an efficient process. These results provide new insight into the physical origin of observed multiphase galactic outflows, and inform galaxy formation models that include coarser treatments of galactic winds. Our results confirm that cool gas observed in outflows at large distances from the galaxy ($\gtrsim1$kpc) likely does not originate through the entrainment of cold material near the central starburst.
\end{abstract}


\section{Introduction}\label{sec:introduction}

Star-forming galaxies commonly feature a multiphase galactic wind, observed at a wide variety of densities, temperatures, and velocities \citep[e.g.][]{Lehnert96, Martin05, Rupke05, Strickland07, Tripp11, Rubin14}, and over a large range of redshifts \citep[e.g.][]{Weiner09, Coil11, Nestor11, Bouche12, Kornei12, Bordoloi16}. Despite their ubiquity, fully characterizing these winds can prove difficult. Spatially-resolved observations of the wind's many phases remain challenging, even for the nearest star-forming systems \citep{Shopbell98, Westmoquette09, Rich10, Leroy15}. Different observational techniques and instruments are required for different phases, so amassing a complete picture for even a single galaxy represents a large coordinated effort. At higher redshifts, absorption line studies that trace outflowing gas in and around star-forming galaxies can be challenging to interpret as they require making assumptions about the wind's geometry \citep[e.g.][]{Rubin11, Bouche12}. While much progress has been made in recent years thanks to the installation of the Cosmic Origins Spectrograph on the \textit{Hubble Space Telescope}, large uncertainties still exist regarding the contributions of different phases of winds to the net mass, momentum, and energy content of outflows \citep{Heckman15}.

Winds also play an important role in theoretical studies of galaxy evolution. Supernova-driven winds provide an
attractive method of feedback in cosmological simulations, allowing galaxies to regulate their star formation rates and gas supply over cosmic time \citep[e.g.,][]{Oppenheimer08, Dave11, FaucherGiguere11, DallaVecchia12, Muratov15}. 
Recent simulations have successfully reproduced the galaxy stellar mass function across a wide range of redshifts by including phenomenologically-motivated wind models \citep{Vogelsberger14, Schaye15, Dave16}. 
However, the processes that launch winds and govern their evolution as they escape galaxies remain unresolved on the scale of cosmological simulations. We currently must turn to smaller-scale, higher-resolution 
simulations to learn more about the physical nature of the winds themselves.

On these smaller physical scales, idealized simulations of galactic winds have also presented a theoretical challenge. Both analytic studies and hydrodynamic simulations of winds have had difficulty accelerating cool gas to the velocities observed in winds, because the dense phases get destroyed by hydrodynamic instabilities too quickly \citep[e.g.,][]{Zhang15, Scannapieco15, Bruggen16}. Magnetic fields may play an important role in stabilizing the cool gas \citep{McCourt15}, but without realistic comparisons to observations the most important physical processes at play in multiphase winds are difficult to ascertain. A detailed analysis of the momentum and energy budget of gas in different phases in these hydrodynamic simulations has not yet been conducted. This data would be valuable both for improving sub-grid prescriptions of winds in cosmological simulations, and for comparing with observations to better determine where our theoretical understanding of winds fails. However, such a study requires high resolution across a large simulation volume in order to track the gas in different phases for significant periods of time.

In this work, we aim to improve our theoretical understanding of multiphase galactic winds via high resolution, idealized simulations. Using the recently released Graphics Processor Unit (GPU)-based code \cholla\footnote{A public version of the \cholla code is available at: http://github.com/cholla-hydro/cholla} \citep{Schneider15}, we can perform hydrodynamic simulations of the interaction between cool and hot phases of a starburst-driven wind at high resolution ($<0.1$pc) over a large volume ($>100$pc). The code performs well enough to compute such simulations on a static mesh, and thus capture the interaction between the different phases of gas across a much larger region than any previous study \citep[e.g.][]{Cooper09, Scannapieco15, Banda-Barragan16}. The ability to track gas in each phase over long periods of time allows a direct probe of the momentum coupling between the hot and cool phases of the wind. In addition, the calculations add an element of physical realism to the cool gas by changing the initial density structure of the multiphase clouds to better match the features seen in spatially-resolved outflows of dense gas.

Our simulations model a multiphase galactic wind as cold, dense interstellar or circumgalactic medium clouds embedded within a hot, rarified background flow driven by supernovae. Because the cool material starts at rest with respect to the background wind, the initial interaction between the two phases drives a shock into the dense cloud. While the current work focuses on the cloud densities, shock mach numbers, and physical scales relevant to galactic winds, the adopted numerical setup allows for comparisons with previous investigations of cloud-shock interactions.

Because of its ubiquity in the ISM, the shock-cloud interaction problem has been studied by many authors. Early numerical work by \cite{Klein94} investigated the case of a planar shock interacting with a spherical cloud using two-dimensional, adiabatic simulations. Their work indicated that clouds encountering a shock typically survive for a few ``cloud crushing times," roughly the timescale for the initial shock to propagate through the cloud. For strong shocks, the cloud crushing time depends on the density contrast between the cloud and the ambient medium, the size of the cloud, and the speed of the shock. Earlier under-resolved numerical work came to similar conclusions \citep{Bedogni90, Nittmann82}. These studies found that shocked clouds travel $\sim8$ cloud radii before mixing with the ambient medium as a result of hydrodynamic instabilities. Adiabatic three-dimensional simulations \citep{Stone92, Xu95} corroborated the two-dimensional results, and additionally attempted to account for different cloud geometries. Cloud geometry and orientation in those simulations did not affect the timescale for cloud fragmentation, but did substantially affect the late-time morphology of the clouds before they were destroyed.

These early studies could reasonably ignore radiative cooling effects by limiting their studies to small clouds.
In larger scale problems where the cooling timescale is smaller than the dynamical timescale, thermal energy losses must be included. Many authors have investigated this regime \citep[e.g.,][]{Mellema02, Fragile04, Melioli05, Cooper09}, and demonstrated that radiative cooling inhibits destruction of the dense material and extends the lifetime of the cloud relative to the adiabatic case. Rather than efficiently mixing with the hot post-shock wind, radiatively-cooling clouds tend to get strung out into filaments containing individual ``cloudlets" of dense gas that can survive much longer. Other authors have investigated the effects of conduction \cite[e.g.,][]{Marcolini05, Orlando05, Bruggen16, Armillotta16} and magnetic fields \cite[e.g.,][]{MacLow94, Fragile05, Shin08, McCourt15, Banda-Barragan16} on the cloud-shock interaction, with varying results for the stabilization of the cloud.

While multiple previous works studied a range of potentially-important physics, few explored the impact of the initial structure of the cloud on the results of cloud-shock interactions. Early work focused on modeling supernova remnants in the ISM, and a simple spherical cloud provided a sufficient approximation for the initial conditions. In radiatively-cooling galactic winds, however, the initial morphology of the cloud may have a profound effect on its evolution. Only \citet{Cooper09} previously studied how the internal structure might influence the cloud destruction, using a fractal cloud as a proxy for a realistic cloud in a galactic wind. They found that fractal clouds survived for less time than initially spherical clouds. More recently, \citet{Schneider15} examined how a turbulent interior cloud structure can alter the cloud crushing timescale in adiabatic simulations.

Our current study aims to better quantify the differences in the physical picture for inhomogeneous clouds, and more broadly describe the way the gas phases in the outflow evolve. Specifically, we attempt to capture the region of parameter space relevant for the cool ($\sim10^4$ K) clouds observed in galactic winds near the disks of star-forming galaxies. In this regime, the wind can be adequately modeled as a hot ($\sim10^6$ K), supersonic fluid containing a population of embedded clouds of denser, cooler, initially stationary material. Depending on the exact density contrast between the cool and hot phases, the cooling timescale may fall below the local dynamic timescale and the simulations therefore should include radiative cooling. Other potentially relevant effects, such as conduction and magnetic fields, we leave for future study.

An outline of our paper follows. We describe in Section~\ref{sec:wind_model} the model used to study the interaction between the multiple phases of the wind. In Section~\ref{sec:simulations} we explain the setup of our wind simulations. Section~\ref{sec:cloud_evolution} presents the qualitative evolution of the wind-cloud interaction, including the impact of the initial surface density of the cool gas on the cloud evolution. In Section~\ref{sec:phase_structure}, we describe in detail the density and temperature structure of the multiphase outflow. In Section~\ref{sec:momentum_coupling} we study the velocities of the gas and describe how momentum distributes between different phases of the wind. Section~\ref{sec:resolution} presents a resolution study focused on increasingly small-scale features in \turb clouds. Section~\ref{sec:discussion} contains our interpretation of these results, including a discussion of our findings in relation to previous work, possible effects of incorporating additional physical processes, and an analysis of the fate of dense gas within a gravitational potential. We summarize in Section~\ref{sec:summary}.

\section{A Multi-Component Wind Model}\label{sec:wind_model}

Theories of starburst galaxies have long suggested that the combination of stellar winds and supernovae should drive a hot ($\sim 10^8$ K) wind out of the starburst region \citep[e.g.][]{Chevalier85}. This hot wind fluid remains difficult to observe directly, requiring high spatial resolution X-ray spectra. In nearby the starburst galaxy M82 where such observations are possible, the detection of a diffuse $\sim4\times10^7$ K plasma in the central region indicates the presence of a hot wind \citep[e.g.][]{Griffiths00, Strickland07}. Our current study assumes that stellar feedback can drive such a wind, and that the coupling of energy and momentum from the hot wind fluid with cooler gas leads to the multiphase winds seen in many starburst systems \citep[see the review by][]{Veilleux05}. In this work, we seek to better quantify the coupling between the hot, rarified phase of galactic winds, and the cooler, denser outflowing gas that is nearly ubiquitously observed in rapidly star-forming systems.

We attempt to account for the origin of a multiphase wind by modeling the interaction between a hot fast outflow and the cold, dense ``clouds" it may entrain. Adequately capturing the hydrodynamic processes occurring in such a scenario requires resolutions of $\approx2$ orders of magnitude smaller than the size of the cool clouds in question. Even with a tool like \chollans, modeling clouds with sizes of $\sim10$ pc requires an idealized set of simulations to probe sufficiently the interactions between the hot and cool gas phases in a wind. Correspondingly our simulations examine dense clouds in a box with a background wind (see Figure~\ref{fig:initial_conditions}), representing cool material exposed to the hot phase of a wind. In the following two subsections, we detail our models for both the hot background and cool cloud components of these multiphase winds.

\subsection{Hot Wind Component}\label{sec:hot_wind}

We seek to model the impact of a hot, supernovae driven outflow on cooler material. Given a small set of assumptions including spherical symmetry and negligible radiative cooling, the hot phase of the wind can be modeled analytically at distances close to the plane of a galaxy. All of our simulations use an analytic model of the hot wind as a constant background state with properties set using the adiabatic wind model of \citet[][hereafter, CC85]{Chevalier85}. The CC85 model envisions a hot wind driven by central energy and mass input from stellar feedback processes. Three input parameters determine the solutions to the model: the energy input as a function of time, $\dot{E}$, the mass input as a function of time, $\dot{M}$, and the size of the driving region within which energy and mass are injected, $R_{*}$. By $\dot{M}$ we mean the total mass injection rate to the wind due to supernovae and mass-loading, \textit{not} the star formation rate. With these parameters, a solution to the set of spherical hydrodynamic equations can be found that smoothly transitions from subsonic within the driving region, to supersonic at further radii. The solutions cross the sonic point at $r = R_{*}$.

\begin{figure}
\includegraphics[width=1.0\linewidth]{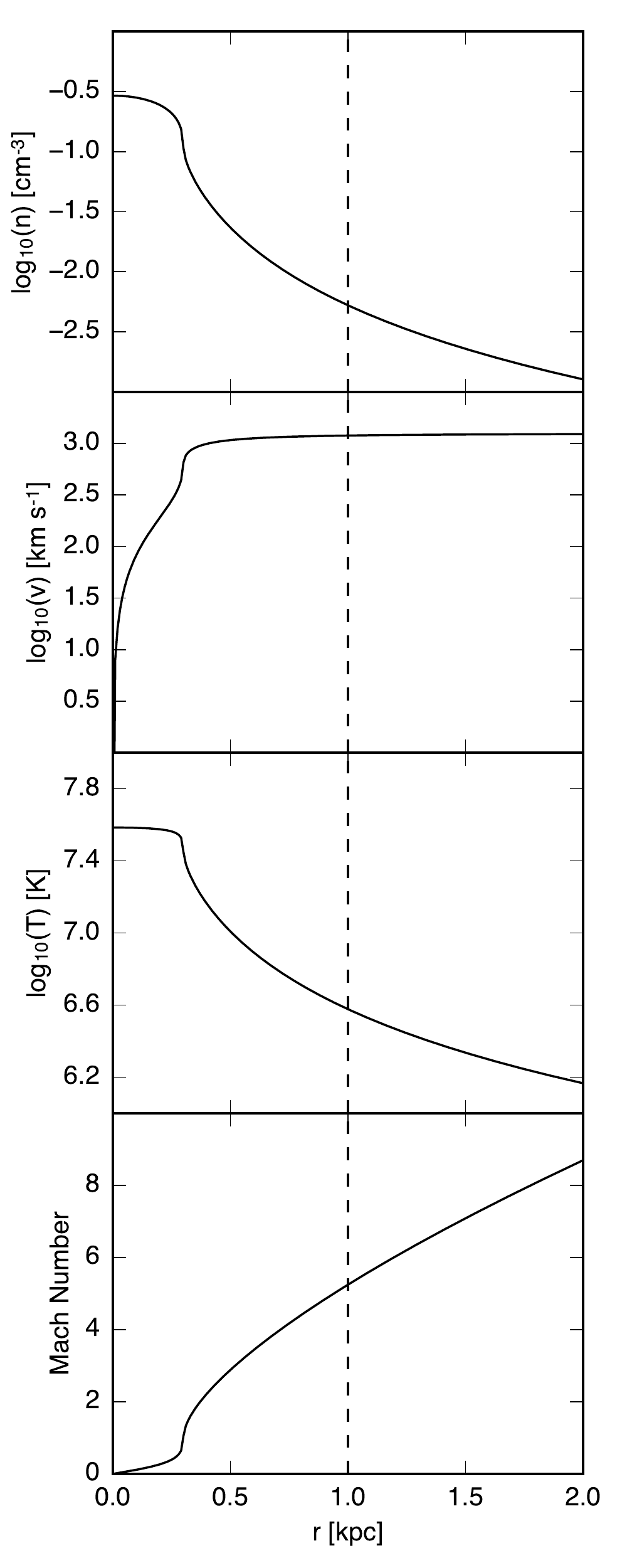}
\caption{The adiabatic wind model used in all simulations. The top three panels display physical values of number density, radial velocity, and temperature as a function of radius. The fourth panel shows the dimensionless mach number of the wind, which crosses the sonic point at $r = R_{*} = 300$ pc in our model. The wind-cloud simulations use values at $r = 1$~kpc for the background wind, shown with the dashed vertical line in each panel.}
\label{fig:wind_model}
\end{figure}

We choose the input parameters for our version of the CC85 model according to the fits derived by \citet{Strickland09} using \textit{Chandra} X-ray observations of the nearby starburst galaxy M82. In particular, we set $\dot{E} = 10^{42}$ erg $\mathrm{s}^{-1}$, $\dot{M} = 2$ $\mathrm{M}_{\odot}$ $\mathrm{yr}^{-1}$, and $R_{*} = 300$ pc. In interpreting their results, we have made additional assumptions about the wind mass-loading factor, $\beta$ and the supernova thermalization fraction $\alpha$ for which they give a range of correlated values. Here we are using the \citet{Strickland09} interpretation of $\beta$ meaning the fraction of total mass injected into the wind as compared to the mass injected by supernovae and stellar winds, $\beta = \dot{M} / \dot{M}_\mathrm{SN+SW}$. Likewise, our definition of $\alpha$ corresponds to their $\epsilon$, and refers to the fraction of the supernova energy that is deposited in low-density gas and does not suffer large radiative losses before being incorporated into the wind. We take $\beta = 1.42$ and $\alpha = 0.33$, values near the middle of the acceptable range of fits reported by \citet{Strickland09}. These choices give a central temperature of $T_{c} \gtrsim 10^{7.5}$ K in the driving region, consistent with estimates made from the X-ray emission \citep[see][Table 2]{Strickland09}. The resulting values for number density, velocity, temperature, and mach number as a function of radius for this model are displayed in Figure~\ref{fig:wind_model}.

For the background flow in our multiphase wind simulations, we use the physical parameters of the CC85 wind model shown in Figure~\ref{fig:wind_model} at $r = 1$ kpc. These values are
\begin{itemize}
\item[] $n_\mathrm{wind} = 5.2626\times10^{-3}$ cm$^{-3}$,
\item[] $v_\mathrm{wind} = 1.1962\times10^{3}$ km s$^{-1}$ = $1.2225$ pc kyr$^{-1}$,
\item[] $P_\mathrm{wind}/k = 1.9881\times10^{4}$ cm$^{-3}$ K,
\end{itemize}
where $k$ is the Boltzmann constant. At $r=1$ kpc, the wind pressure corresponds to a temperature of $T_\mathrm{wind} = 3.7778\times10^{6}$ K, as displayed in Figure~\ref{fig:wind_model}. We choose a radius of 1 kpc to set the background wind properties in our simulations for several reasons. First, we wish to capture the mass-loading of the wind outside the driving region, which restricts us to hot wind properties at radii $r>300$ pc. Second, we will model the interactions between cool and hot material in a simulation volume with a physical length of 160 pc. Given that the wind properties in our simulations remain approximately constant across this volume and the hot wind density, temperature, and mach number changes most rapidly just outside the driving region (see Figure~\ref{fig:wind_model}), we favored an initial radius of $r\sim1$ kpc over smaller radii. Further, the best observations of a multiphase wind come from M82, where cool material clearly resides at  $r>1$ kpc above the disk \citep{Leroy15}. The fits derived by \citet{Strickland09} suggest that the hot wind should not yet have suffered serious radiative losses at $r\sim1$kpc, which would invalidate the CC85 model \citep{Zhang15, Thompson16}.

\subsection{Cool Cloud Component}\label{sec:cool_clouds}

\begin{figure*}
\includegraphics[width=1.0\linewidth]{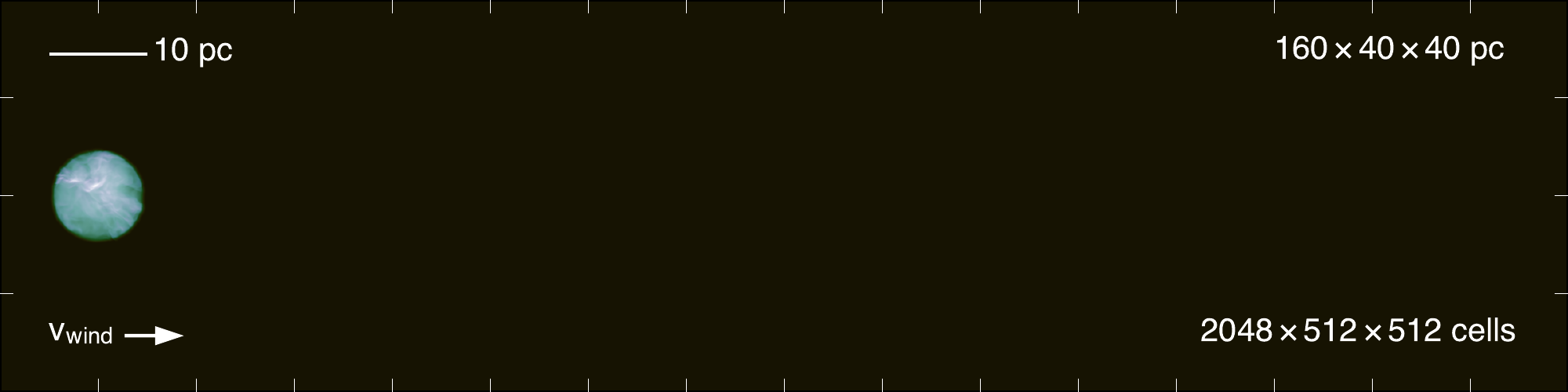}
\caption{The initial conditions for one of our cloud-wind simulations. Each simulation box is much larger than the initial size of the cloud, so we can track the long-term evolution of cloud material, even after it has been stripped from the main body of the cloud. This density projection shows the $\tilde{n} = 1$ $\mathrm{cm}^{-3}$ \turb cloud. The initial density distribution for the cloud material in this simulation is displayed in Figure~\ref{fig:cloud_pdf}.}
\label{fig:initial_conditions}
\end{figure*}

To capture the multiphase nature of galactic winds, our simulations also include a cool component representing interstellar or circumgalactic material. As with previous studies of galactic winds, we model this second component as dense clouds initially at rest with respect to the hot wind \cite[e.g.][]{Scannapieco15, Bruggen16}. Our aim is to extend previous studies by examining the detailed momentum and energy coupling between different wind phases, so we consider both idealized spherical clouds and more realistic \turb clouds with a distribution of interior densities set by turbulent processes. Observations have revealed cool gas in outflows at a variety of densities and temperatures (see references in Section~\ref{sec:introduction}). By varying the median density of the cold gas, $\tilde{n}$, and its interior density structure, we are able to model a range of properties for the dense component of the wind. Both the median number density and the cloud morphology affect the integrated surface density of the cold gas, $\Sigma_\mathrm{cl} = M_\mathrm{cl} / \pi R_\mathrm{cl}^2$, where $M_\mathrm{cl}$ is the total mass of the cloud and $R_\mathrm{cl}$ is the cloud radius. The surface density influences how momentum transfers from the hot wind to the cold component, as discussed below.

As the hot wind destroys the clouds, cool material will both heat and rarify. This material will accelerate as momentum transfers from the hot wind, and enter the outflowing wind with a range of velocities. To adequately track all of this material, we perform simulations using boxes with long aspect ratios. The simulation boxes feature transverse dimensions equal to 8 $R_\mathrm{cl}$, and a long dimension along the wind direction of 32 $R_\mathrm{cl}$. Figure~\ref{fig:initial_conditions} displays an example of an entire box, showing a density projection of the $\tilde{n} = 1$ $\mathrm{cm}^{-3}$ \turb cloud at time $t = 0$. The clouds are initially centered 2 $R_\mathrm{cl}$ from the left boundary of the box to capture the resulting bow shock. The long dimension of the box enables the simulations to follow the bulk of the cloud as it accelerates \textit{and} track material stripped from the main cloud body.

\subsubsection{Cloud and Sphere Models}

\begin{figure}
\includegraphics[width=1.0\linewidth]{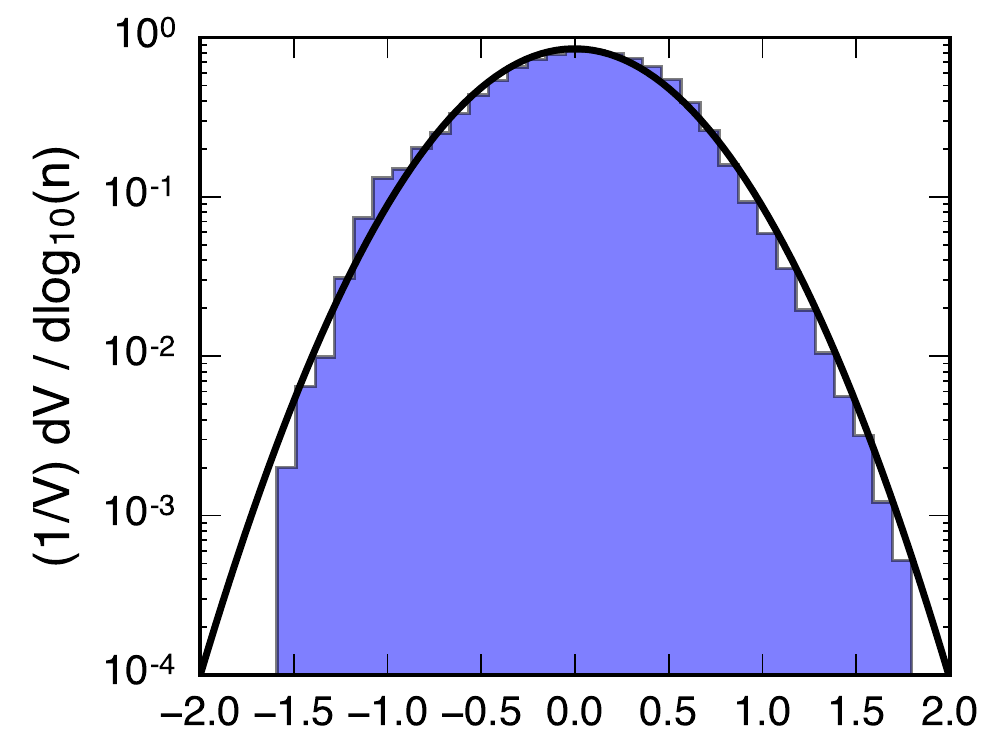}
\caption{A normalized histogram of the initial density distribution for the $\tilde{n} = 1$ \turb cloud is shown in blue, fit with a gaussian in log-space. Note: this distribution does not include the density taper at the edges of the cloud, as its purpose is to illustrate the lognormal distribution of the bulk of the cloud's mass.}
\label{fig:cloud_pdf}
\end{figure}

Each cloud initially sits at rest and in thermal pressure equilibrium with the surrounding hot wind. The clouds in our study are not dense enough to be gravitationally confined, allowing us to neglect self-gravity. The density of the cloud material therefore determines its initial temperature. We initialize the spherical clouds with constant interior temperature, and initialize the interior temperatures of regions within the \turb clouds along an appropriate isobar. We list the temperatures and median and mean densities of the cloud initial conditions in Table~\ref{tab:simulations}. We list the full range of temperatures for the \turb clouds - their median temperature matches the spheres. For both the spheres and \turb clouds we taper the densities at the edge, starting at a radius of 4.5 pc, such that the cloud density smoothly transitions from the median to the wind density. The density taper has the form
\begin{equation}
n(r)_\mathrm{cl} = \tilde{n}\exp[(\mathrm{ln}(n_\mathrm{wind}/ \tilde{n})/(R_\mathrm{cl} - 4.5)] |r- 4.5|,
\end{equation}
where $r$ is the distance from the center of the cloud, and $\tilde{n}$ is the median cloud density. $R_\mathrm{cl} = 5$~pc for all of our clouds. To create the \turb clouds, we excise a region from a Mach 5 isothermal turbulence simulation \citep{Robertson12}, and scale the density linearly to match the desired median. Figure~\ref{fig:cloud_pdf} shows a histogram of the initial gas densities for the $\tilde{n} = 1$ $\mathrm{cm}^{-3}$ \turb cloud simulation.

\subsubsection{Cloud Crushing Time}

To interpret our results in relation to previous studies of cloud-shock interactions, we make use of the ``cloud-crushing time" $t_\mathrm{cc}$ initially devised by \cite{Klein94}. When the hot wind first interacts with the cool cloud, a shock drives into the overdense gas (and a reverse shock reflects into the oncoming wind). The cloud-crushing time estimates how long the initial shock takes to pass through and compress the cloud. This timescale can be calculated in terms of known quantities by relating the initial density contrast of the cloud to the wind $\chi = n_\mathrm{cl} / n_\mathrm{wind}$ along with the radius of the cloud $R_\mathrm{cl}$ and the wind velocity $v_\mathrm{wind}$ as
\begin{equation}
t_\mathrm{cc} = R_\mathrm{cl} \chi^{\frac{1}{2}} / v_\mathrm{wind}.
\label{eqn:t_cc}
\end{equation}
We use the cloud-crushing time as an evolutionary timescale for the clouds in our simulations, and report $t_\mathrm{cc}$ for each of our simulations in Table~\ref{tab:simulations}. The derivation leading to Equation~\ref{eqn:t_cc} assumes that the pre-shocked gas in the cloud evolves adiabatically, allowing for an estimate of the shock speed within the cloud from the density contrast and wind speed. In a radiatively-cooling cloud the pre-shock conditions change as the cloud cools, decreasing the sound speed and slowing the shock. The cloud crushing times listed in Table~\ref{tab:simulations} therefore represent imperfect estimates of the cloud compression timescale, and our simulations show they underestimate the duration of this phase. In our simulations, maximum compression is typically reached at $\sim 2$ \tccns.

\subsubsection{Cloud Cooling Time}

The cooling time provides another important evolutionary timescale for the clouds we simulate. If the cloud cooling time greatly exceeds the cloud crushing time, we expect the clouds to evolve similarly to the well-studied adiabatic case \citep{Klein94, Xu95, Poludnenko02, Nakamura06}. If the cloud crushing time grows much longer than the cooling time, the radiative loss of energy may affect the cloud properties substantially before the initial shock can disrupt it. We can estimate the cooling time of the clouds in our simulation as their thermal energy divided by the rate at which energy is lost owing to radiative cooling as 
\begin{equation}
t_\mathrm{cool} = \frac{3 n_\mathrm{cl} k T_\mathrm{cl}}{2 \Lambda(T_\mathrm{cl})},
\label{eqn:t_cool}
\end{equation}
where $k$ is the Boltzmann constant and $\Lambda(T)$ is the cooling rate in erg $\mathrm{s}^{-1}$ $\mathrm{cm}^{-3}$, evaluated at the cloud temperature (see Appendix~\ref{app:cooling} for information on our calculation of cooling rates). In theory, we might like to use $T_\mathrm{cl}$ and $n_{cl}$ post-shock to calculate the cooling time. Given the complex nature of the cooling function, these quantities often prove impossible to calculate analytically \citep{Creasey11}. Additionally, the lognormal gas distribution in our turbulent clouds means that $n$ and $T$ vary for different parts of the cloud. Instead, we calculate the cloud cooling times using the median pre-shock density and temperature to compare most directly with the cloud-crushing time. These cooling times appear in Table~\ref{tab:simulations}. Our simulations sample the transition from clouds dominated by adiabatic evolution to those in the radiatively cooling regime. As discussed in Section~\ref{sec:cloud_evolution}, our simulations reproduce this transition and thereby justify our assumptions used to compute the cooling time.

\section{Simulations}\label{sec:simulations}

\afterpage{
\begin{deluxetable*}{cccccccccc}
\tabletypesize{\scriptsize}
\tablecaption{Summary of simulations performed, and physical parameters of interest for the cloud in each simulation.\label{tab:simulations}}
\tablehead{
\colhead{Model} &
\colhead{$\tilde{n}_\mathrm{cl}$\tablenotemark{a}} &
\colhead{$\bar{n}_\mathrm{cl}$\tablenotemark{b}} &
\colhead{$\Sigma_\mathrm{cl}$\tablenotemark{c}} &
\colhead{$N_{H}$\tablenotemark{d}} &
\colhead{$T_\mathrm{cl}$\tablenotemark{e}} & 
\colhead{$M_\mathrm{cl}$\tablenotemark{f}} & 
\colhead{$\chi$\tablenotemark{g}} & 
\colhead{$t_\mathrm{cc}$\tablenotemark{h}} & 
\colhead{$t_\mathrm{cool}$\tablenotemark{i}} \\
\colhead{Name} &
\colhead{[cm$^{-3}$]} &
\colhead{[cm$^{-3}$]} &
\colhead{[$M_{\odot}$ pc$^{-2}$]} &
\colhead{[cm$^{-2}$]} &
\colhead{[K]} &
\colhead{[M$_{\odot}$]} &
\colhead{} &
\colhead{[kyr]} &
\colhead{[kyr]}
}
\startdata
S01 & 0.1 & 0.086 & 0.013 & $2.87\times10^{18}$ & $1.988\times10^{5}$ & 1.05 & $1.9\times10^{1}$ & 17.8 & 26.0 \\
S05 & 0.5 & 0.45 & 0.064 & $1.43\times10^{19}$ & $3.976\times10^{4}$ & 5.04 & $9.5\times10^{1}$ & 39.8 & 3.96 \\
S1 & 1.0 & 0.88 & 0.13 & $2.85\times10^{19}$ & $1.988\times10^{4}$ & 10.0 & $1.9\times10^{2}$ & 56.4 & 1.30 \\
T01 & 0.1 & 0.16 & 0.022 & $2.05\times10^{19}$ & $2.2\times10^{3}$ & 1.73 & $1.9\times10^{1}$ & 17.8 & 26.0 \\
T05 & 0.5 & 0.83 & 0.11 & $1.02\times10^{20}$ & $4.4\times10^{2}$ & 8.61 & $9.5\times10^{1}$ & 39.8 & 3.96 \\
T1 & 1.0 & 1.7 & 0.24 & $2.23\times10^{20}$ & $2.0\times10^{2}$ & 18.8 & $1.9\times10^{2}$ & 56.4 & 1.30 \\
\enddata
\tablenotetext{a}{Median density of the cloud, calculated including all material with a density greater than 1/10 $\tilde{n}$.}
\tablenotetext{b}{Mean density of the cloud, calculated including all material with a density greater than 1/10 $\tilde{n}$.}
\tablenotetext{c}{Surface density of the cloud, calculated as $M_\mathrm{cl} / (\pi R_\mathrm{cl}^2)$.}
\tablenotetext{d}{Maximum initial column density sight-line through the cloud (excluding the hot wind).}
\tablenotetext{e}{Initial temperature of the cloud material, in pressure equilibrium with the wind. The median temperature is provided for the spherical clouds, and the temperature of the highest density regions is provided for the \turb clouds.}
\tablenotetext{f}{Initial mass of the cloud, calculated including all material with a density greater than 1/10 $\tilde{n}$.}
\tablenotetext{g}{Initial density contrast of the cloud with the background wind, calculated using $\chi = \tilde{n}_\mathrm{cl} / n_\mathrm{wind}$.}
\tablenotetext{h}{Cloud crushing time, calculated using Equation~\ref{eqn:t_cc} with the median density and a cloud radius of 5.0 pc.}
\tablenotetext{i}{Cloud cooling time, calculated using Equation~\ref{eqn:t_cool} with the median cloud density and temperature.}
\tablecomments{Simulations with a spherical cloud begin with an 'S'; those with a turbulent cloud begin with a 'T'. All simulations listed in Table~\ref{tab:simulations} are run in a volume with a numerical resolution of $2048\times512\times512$ cells, corresponding to a physical box size of $160\times40\times40$ pc.}
\end{deluxetable*}
}

We ran a total of six ``production" simulations with initial parameters displayed in Table~\ref{tab:simulations}. The first parameter we varied in these simulations was the cloud density distribution. Half the simulations modeled constant density spherical clouds to compare with previous studies. The other half modeled clouds with a lognormal density distribution appropriate for a turbulent gas \citep[e.g.][]{Padoan02, Kritsuk07}. We also varied the median density of the clouds from $\tilde{n} = 0.1$ $\mathrm{cm}^{-3}$ to $\tilde{n} = 1.0$ $\mathrm{cm}^{-3}$ to sample a range of density and temperature phase space. At the low end, this density range samples the transition from adiabatic to radiative evolution of the cool gas. At the higher densities, the evolution of the clouds is very similar when scaled by the cloud-crushing time (See Section~\ref{sec:cloud_evolution}), but higher density clouds have increasingly lengthy lifetimes. As a result, our upper limit on cloud density is set by computational expense.

Throughout this paper, we refer to the production simulations by the names given in Table~\ref{tab:simulations}. Simulations with spherical clouds begin with an `S', and those with turbulent clouds are denoted `T'. The remainder of the name references the median density of the cloud. E.g. the simulation featuring a \turb cloud with a median number density of $\tilde{n} = 1$ $\mathrm{cm}^{-3}$ is `T1'; the spherical cloud with a median number density of $\tilde{n} = 0.1$ $\mathrm{cm}^{-3}$ is `S01', etc. We will sometimes refer to the spherical clouds as ``spheres''; ``cloud'' alone will always refer to a \turb cloud. 

Each of the simulations listed in Table~\ref{tab:simulations} was run in a volume with $N=2048\times512\times512$ cells, and physical dimensions of $160\times40\times40$ pc, yielding a physical resolution for these simulations of $\Delta x =0.07825$ pc/cell. We also carried out a resolution study (see Section~\ref{sec:resolution}), with higher and lower resolution simulations for the $\tilde{n} = 0.5$ \turb cloud. Previous work on this problem has relied on techniques such as adaptive mesh refinement and cloud-tracking (moving the reference frame of the simulation to follow the main body of the cloud) to reduce computational costs. In contrast, we have constant resolution across our entire volume, and as a result we capture the evolution of both low and high density material. Our long boxes also allow us to track material at distances further from the cloud than in previous studies.

We ran our simulations with the \cholla hydrodynamics code \citep{Schneider15}, using piecewise parabolic reconstruction, an HLLC Riemann solver, a simple unsplit integrator, and a dual energy scheme. Optically-thin radiative cooling with a photoionizing UV background was assumed, and implemented using pre-computed Cloudy tables \citep{Ferland13}. Details of the integration scheme, Riemann solver, dual energy, and radiative cooling appear in the Appendices. We used outflow boundaries for all sides of the box except the left $x$-face, where the inflow was set according to the wind properties described in Section~\ref{sec:hot_wind}. All of the production simulations were run for a minimum of 25 $t_\mathrm{cc}$. The ultra high resolution simulation used in our resolution study was only run for 12 \tccns, as it is extremely expensive.

The six simulations listed in Table~\ref{tab:simulations} have high enough resolution to adequately capture the hydrodynamic instabilities that disrupt the spherical clouds, 64 cells/$R_\mathrm{cl}$. These simulations were used to produce the majority of the results in this paper. In addition, we have run low resolution versions of each of the simulations in Table~\ref{tab:simulations} in a volume with half as many cells ($N=1024\times256\times256$). The primary purpose of these low resolution runs was to verify initial conditions, estimate runtimes, and test for convergence. We have also run a low resolution simulation with a different mean molecular weight (see Appendix~\ref{app:cooling}) to test the effect of the cooling implementation on our results.

In total, we have run 14 simulations for this paper. The 7 low resolution simulations were carried out on the \textit{El Gato} cluster at the University of Arizona. The high resolution production simulations and ultra high resolution comparison simulation were carried out on the \textit{Titan} supercomputer at the Oak Ridge Leadership Computing Facility via a director's discretionary time allocation. In sum, these high resolution simulations took $\approx1.5$ million \textit{Titan} core-hours, which corresponds to 50,000 GPU-hours. The time each simulation required varied greatly based on the total length of cloud survival, from the lowest density $\tilde{n} = 0.1$ spherical cloud simulation that required $\approx$~1,500 GPU-hours, to the $\tilde{n} = 1.0$ spherical cloud simulation that required $\approx$~9,400 GPU-hours. The $\tilde{n} = 0.5$ ultra high resolution \turb cloud simulation required $\approx$~14,000 GPU-hours despite only running to 12 \tccns.

\section{Cool Cloud Evolution}\label{sec:cloud_evolution}

We begin our results with a general description of the evolution of the cool gas in the clouds, focusing on the $\tilde{n} = 1$ \turb cloud simulation (T1), our fiducial case. The evolution of cool material when exposed to a hot wind is of interest, given the uncertainty in the theoretical community about whether it is possible to accelerate cool material to the hot wind speed without destroying it \citep[e.g.][]{Scannapieco15, Zhang15}. If cool gas can be efficiently entrained, the process could provide an explanation for the presence of cool material observed at large distances from galaxies \citep[e.g.][]{McCourt15}. The efficiency with which cool material is destroyed will also affect the mass-loading factor of the hot wind, which may in turn affect the chances of thermal instability and rapid cooling of the hot wind \citep{Wang95, Thompson16}. In this section, we describe the overall evolution of the cool gas in our simulations as a function of both cloud morphology and initial surface density. In particular, we focus on the destruction time of the cool material.

The early stages of adiabatic shock-cloud interactions have been thoroughly described in the literature, particularly beginning with the comprehensive study of \cite{Klein94}. Cloud evolution is often described in terms of the cloud-crushing time, $t_\mathrm{cc} = \chi R_\mathrm{cl} / v_\mathrm{wind}$ (see Section~\ref{sec:cool_clouds}). In an adiabatic simulation, the initial compression of the cloud is followed by a downstream expansion, after which the cloud quickly fragments and dense material is destroyed. The entire cloud destruction process typically takes 4 - 5 \tcc for spheres. If the cloud-crushing time is calculated using the median gas density, adiabatic \turb clouds are destroyed in a similar number of crushing times \citep{Schneider15}.

When dense clouds are able to cool radiatively, their evolution follows a different path. The early cloud crushing phase is much more effective as the heat generated by compression is radiated away. In a radiative cloud, gas densities can reach over an order of magnitude above their pre-shock levels. Though radiative clouds still get disrupted within 10 \tccns, the individual dense ``cloudlets" that result can take as long as 40 \tcc to mix into the hot wind \citep{Scannapieco15}.

\subsection{Turbulent Clouds vs Spheres}

\begin{figure*}
\begin{centering}
\includegraphics[width=0.62\linewidth]{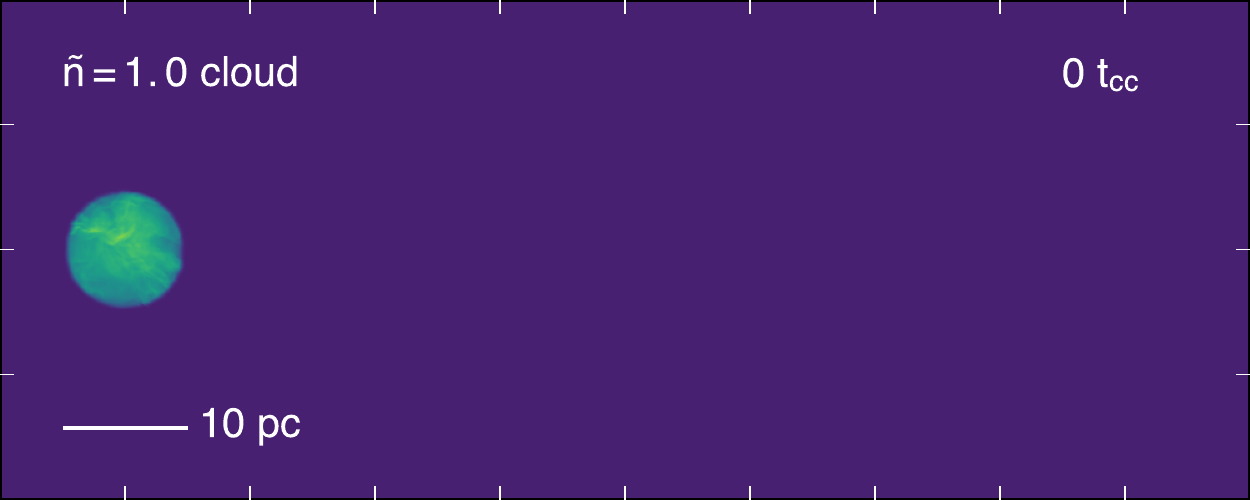}
\includegraphics[width=0.372\linewidth]{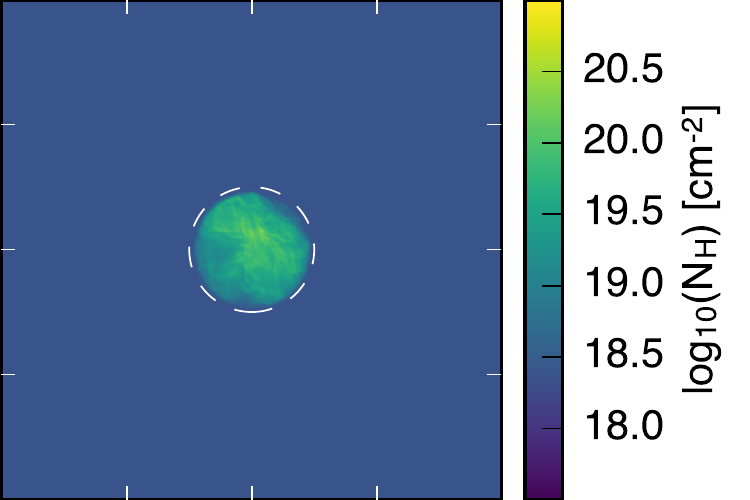}
\includegraphics[width=0.62\linewidth]{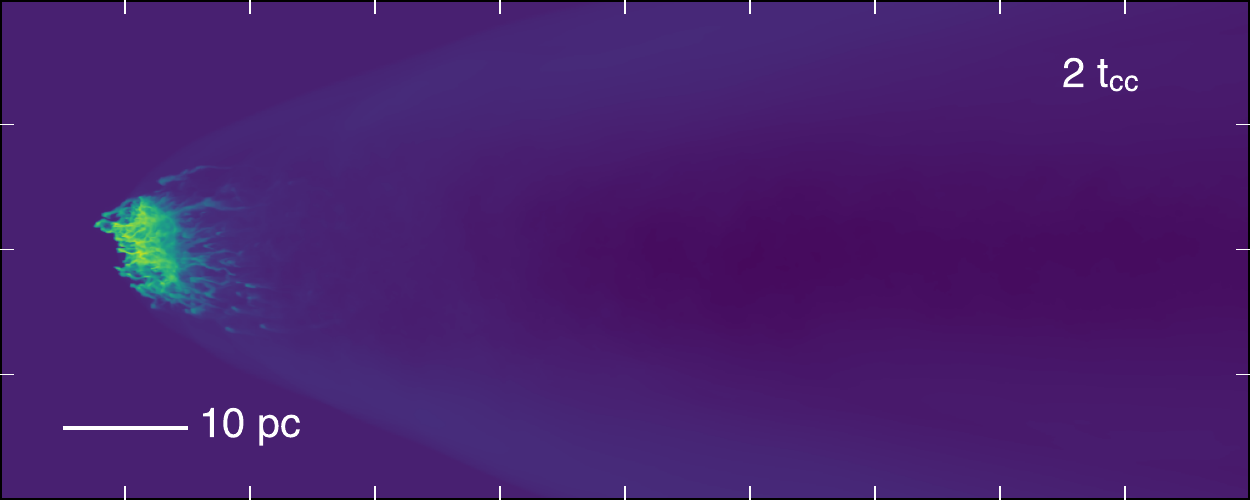}
\includegraphics[width=0.372\linewidth]{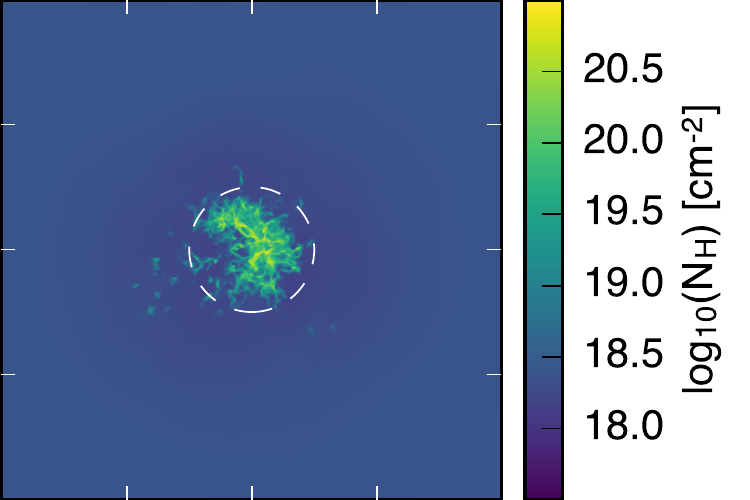}
\includegraphics[width=0.62\linewidth]{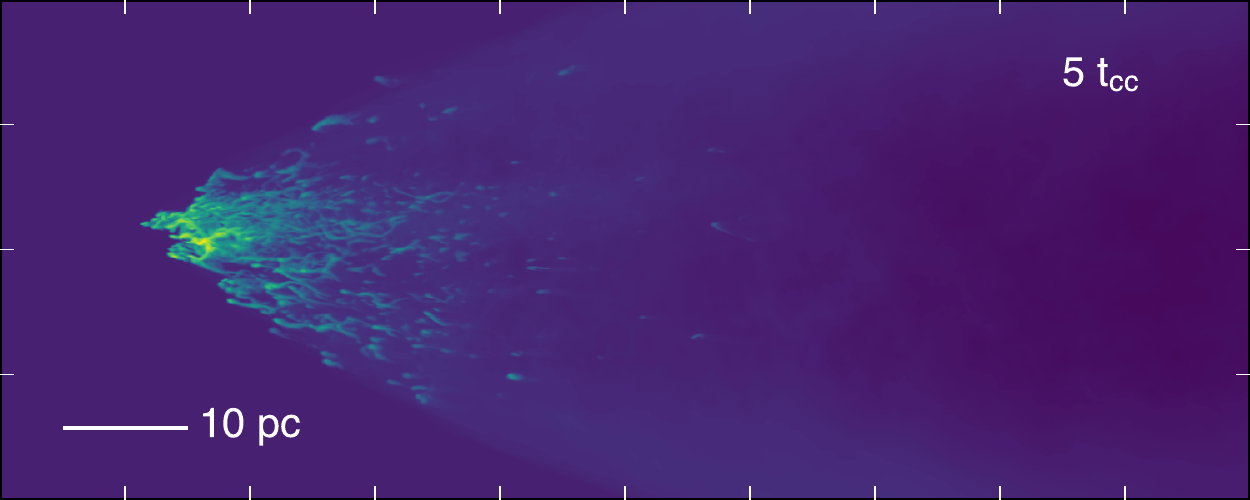}
\includegraphics[width=0.372\linewidth]{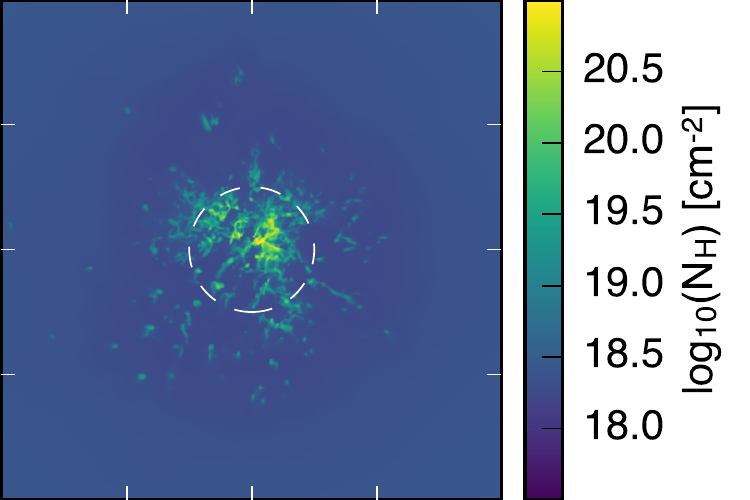}
\includegraphics[width=0.62\linewidth]{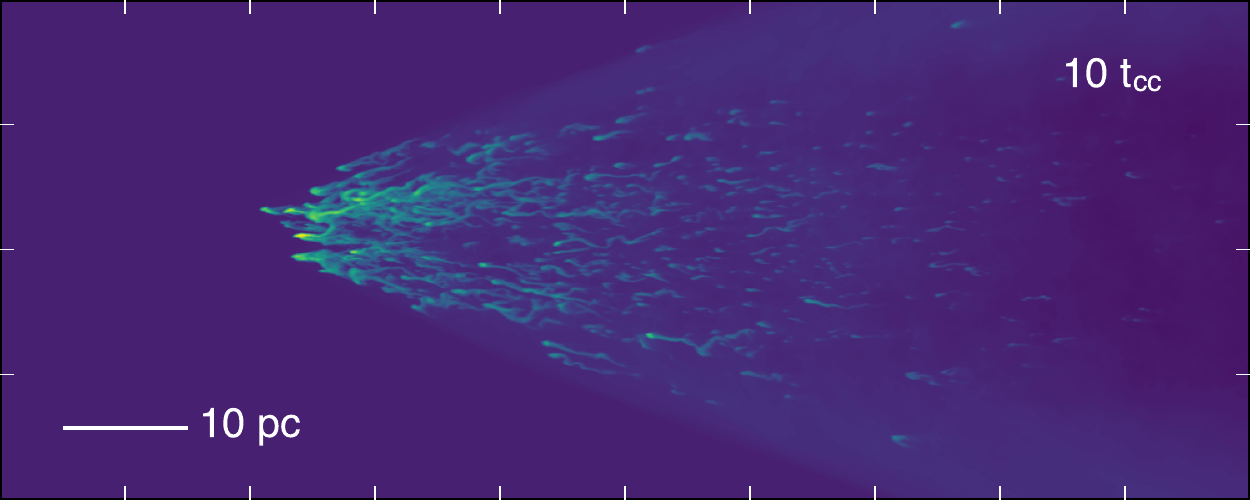}
\includegraphics[width=0.372\linewidth]{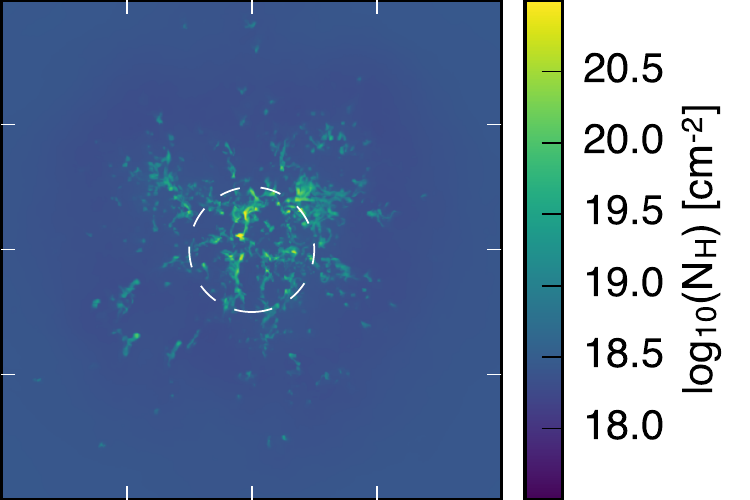}
\includegraphics[width=0.62\linewidth]{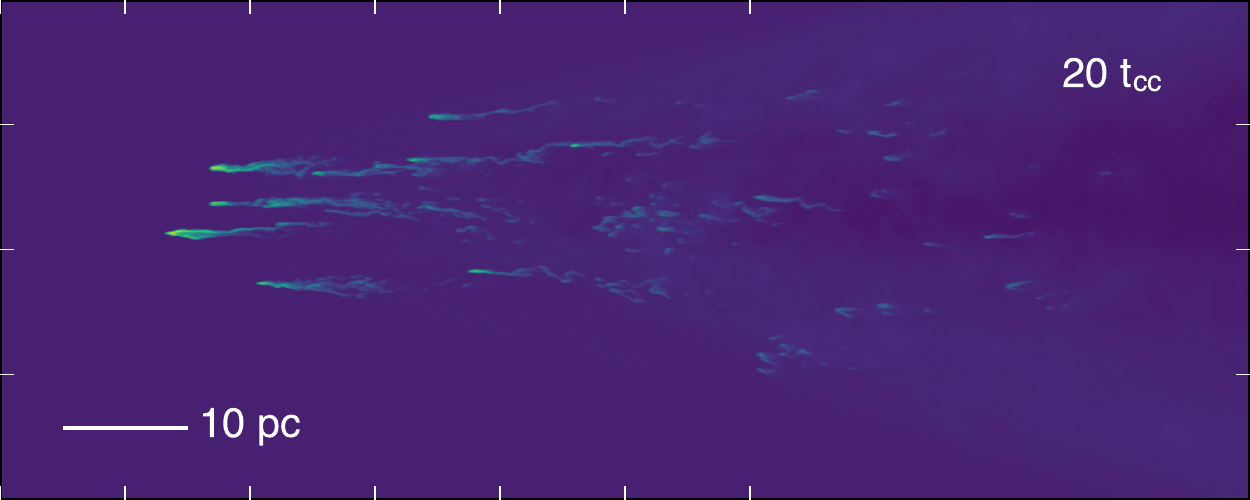}
\includegraphics[width=0.372\linewidth]{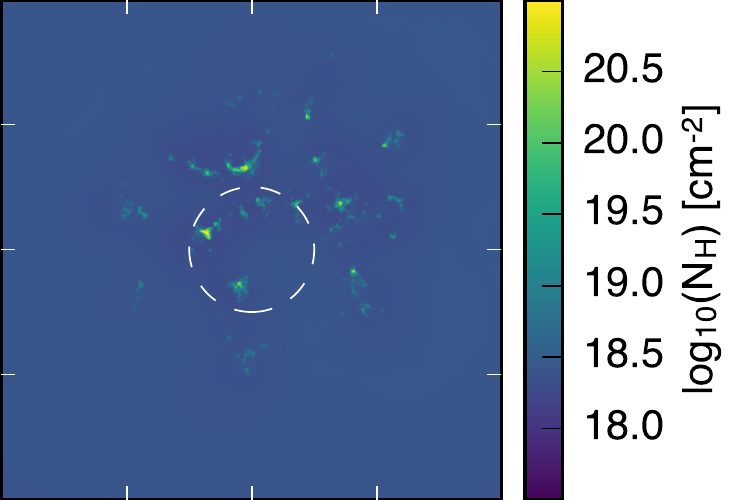}
\caption{Time series evolution of our fiducial simulation, the $\tilde{n} = 1$ $\mathrm{cm}^{-3}$ \turb cloud (model T1). Plots on the left show surface density projected along the $y$-axis, with the wind entering the box from the left, while the right column shows the surface density projected in the direction of the wind velocity. The scale of the axes ticks is 10 pc. Snapshots are shown at $t=0$, 2, 5, 10, and 20 \tccns. The dashed circle in the right column shows the original extent of the cloud. Note: the $y$-projection at $t=20$ \tcc has been shifted by 30 pc to re-center the cloud material.}
\label{fig:cwn1_evolution}
\end{centering}
\end{figure*}

\begin{figure*}
\begin{centering}
\includegraphics[width=0.62\linewidth]{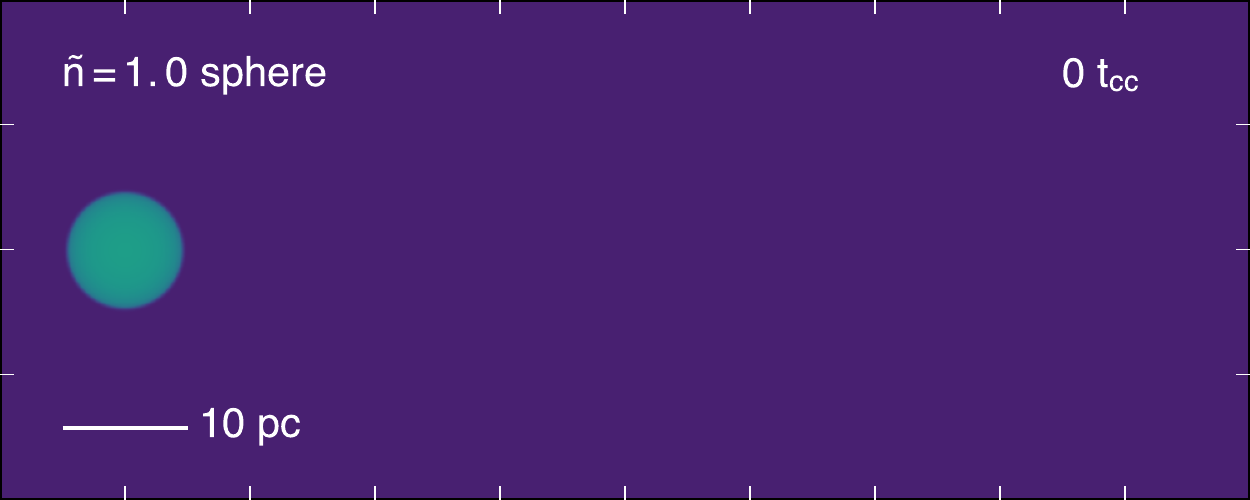}
\includegraphics[width=0.372\linewidth]{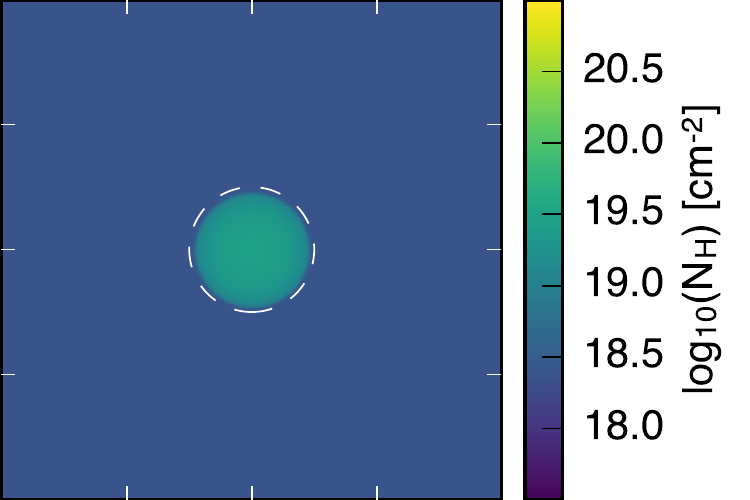}
\includegraphics[width=0.62\linewidth]{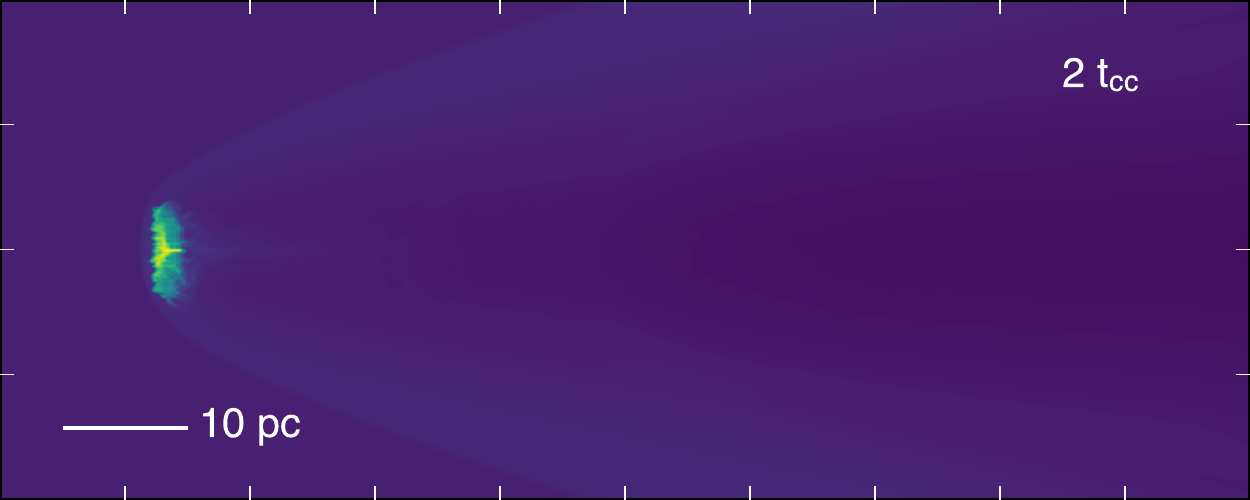}
\includegraphics[width=0.372\linewidth]{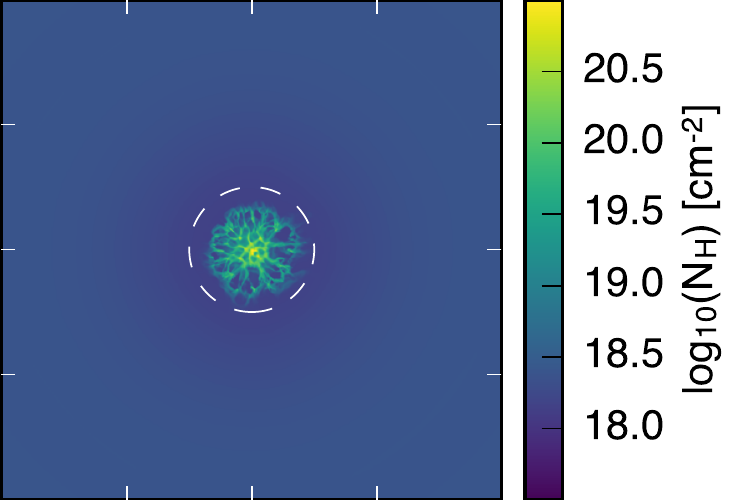}
\includegraphics[width=0.62\linewidth]{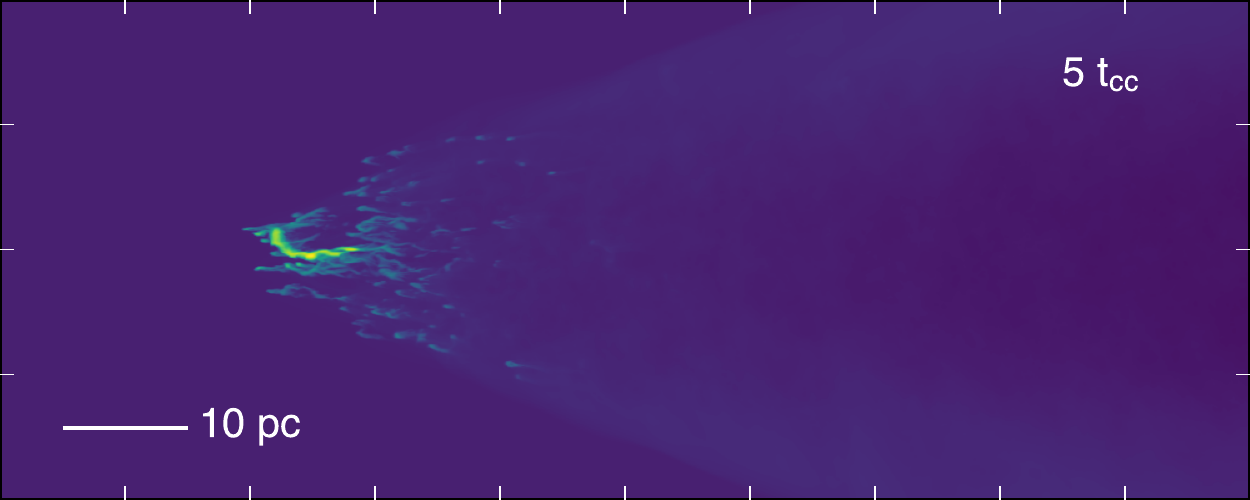}
\includegraphics[width=0.372\linewidth]{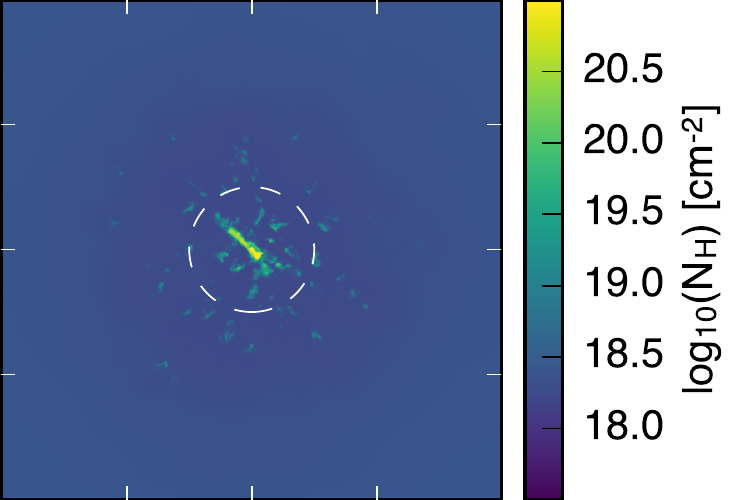}
\includegraphics[width=0.62\linewidth]{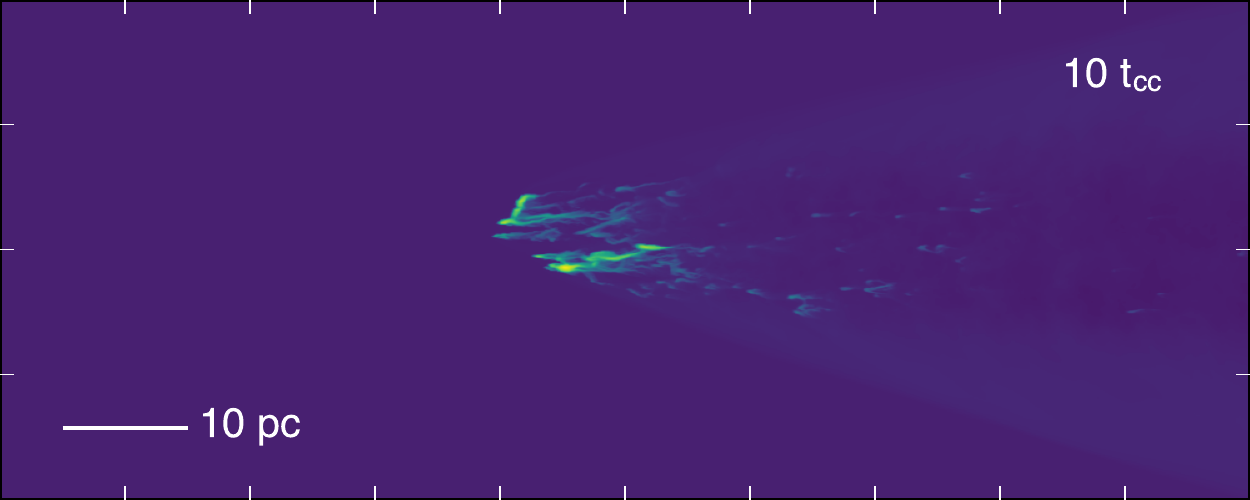}
\includegraphics[width=0.372\linewidth]{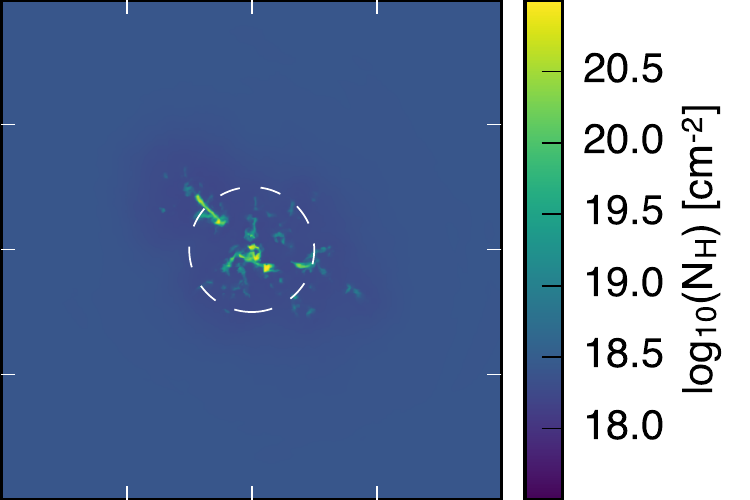}
\includegraphics[width=0.62\linewidth]{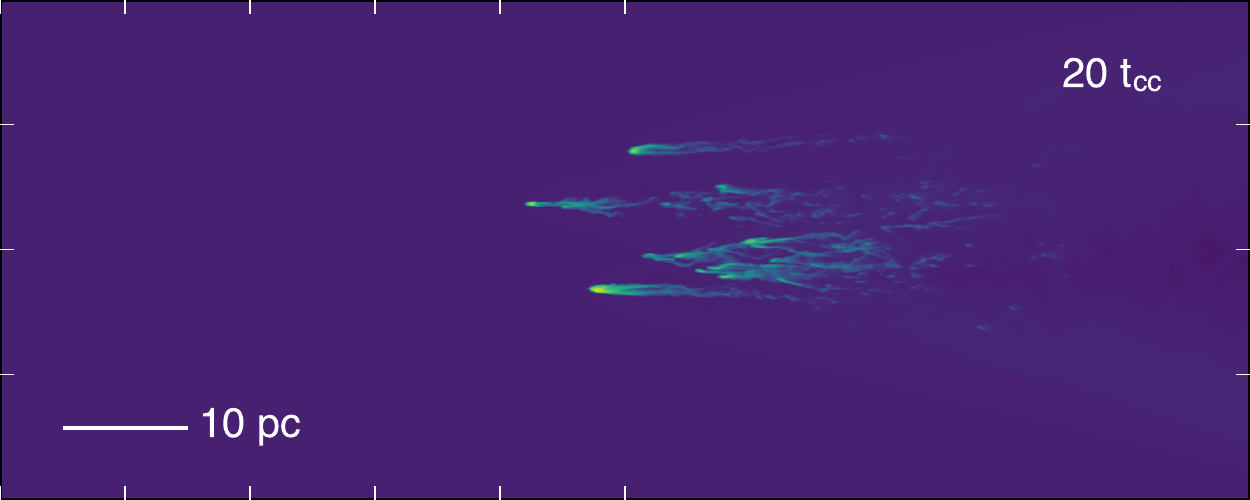}
\includegraphics[width=0.372\linewidth]{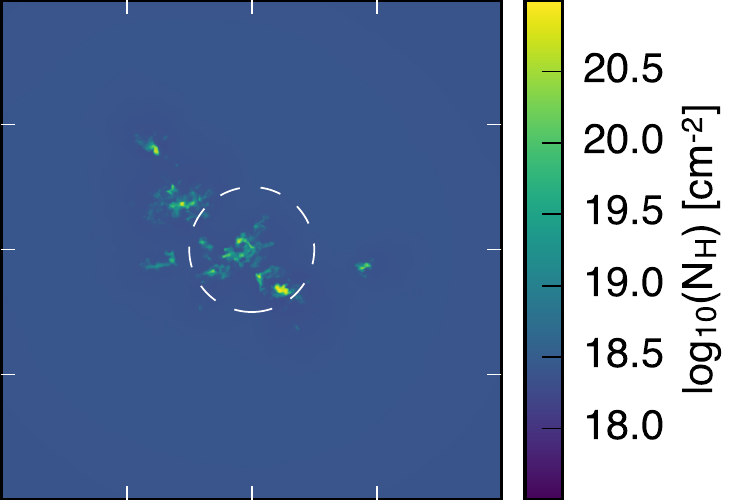}
\caption{Time series evolution of the $\tilde{n} = 1$ $\mathrm{cm}^{-3}$ sphere simulation (model S1). As in Figure~\ref{fig:cwn1_evolution}, the left column shows surface density projected along the $y$-axis, and the right column shows the surface density projected along the wind axis. Snapshots are shown at $t=0$, 2, 5, 10, and 20 \tccns. Note: the reference frame of the $y$-projection at $t=20$ \tcc has been shifted by 40 pc to re-center the cloud.}
\label{fig:swn1_evolution}
\end{centering}
\end{figure*}

In Figures~\ref{fig:cwn1_evolution} and \ref{fig:swn1_evolution} we illustrate the general evolution of the $\tilde{n}$ = 1 $\mathrm{cm}^{-3}$ \turb and spherical cloud simulations (models T1 and S1, Table~\ref{tab:simulations}). 
These figures show the column density of gas in the box (including the hot wind), in both a $y$-axis projection with the hot wind entering the box from the left, and an $x$-axis projection in the direction of the wind. The figures show snapshots of the simulations at 5 representative times - the initial conditions, and after 2, 5, 10, and 20 \tccns.

\begin{figure}
\includegraphics[width=1.0\linewidth]{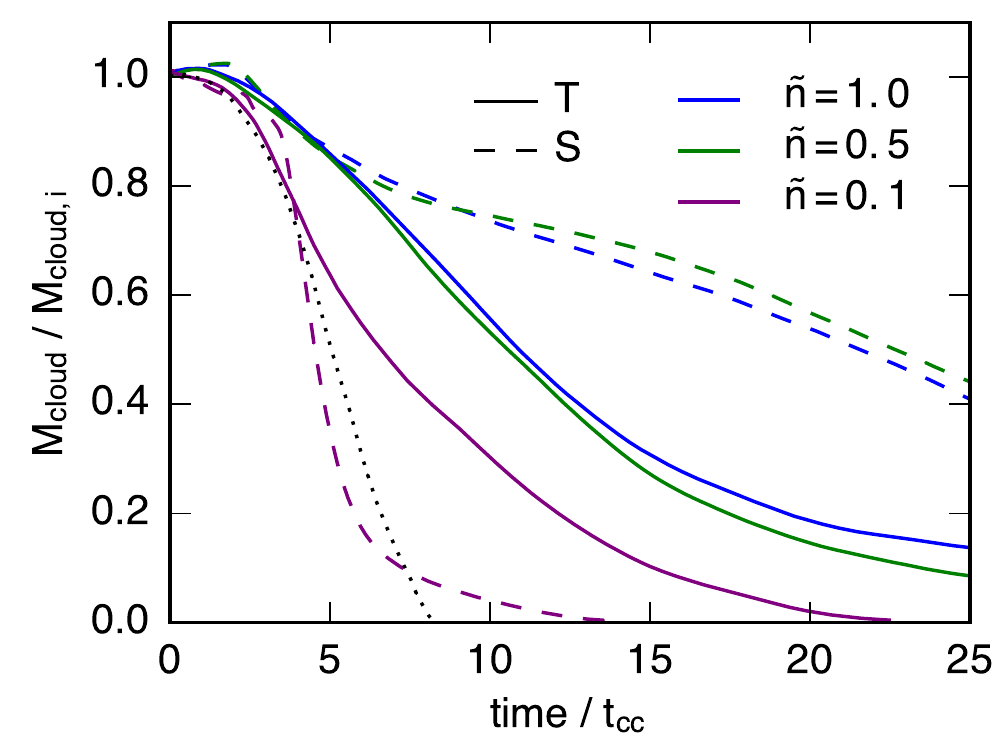}
\caption{The mass evolution (cloud mass divided by initial cloud mass as a function of time) of each of our high resolution simulations. Cloud mass is calculated as the sum of all gas in the simulation with a density greater than 1/3 the initial median density. Tracks for \turb clouds are shown as solid lines, while spherical clouds are shown with dashed lines. The mass-loss track for an $\tilde{n} = 0.1$ adiabatic cloud simulation is also plotted (dotted line) for comparison.}
\label{fig:mass_evolution}
\end{figure}

In both simulations, the maximum average cloud density is reached at $t=2$ \tccns. The maximum average density is $\bar{n} = 5.2$~$\mathrm{cm}^{-3}$ for T1, and $\bar{n} = 6.8$~$\mathrm{cm}^{-3}$ for S1. (We calculate the mean density using material with density greater than 1/10th the initial median density, as in Table~\ref{tab:simulations}.) Despite the similarity of this compression timescale, Figures~\ref{fig:cwn1_evolution} and ~\ref{fig:swn1_evolution} show a drastic difference in cloud morphology at $t=2$ \tccns. While the initial shock has propagated at different speeds through regions of different density for the \turb cloud, the shock has very effectively compressed the sphere into a single flat pancake. As a result, the evolution of the two types of clouds diverges strongly at later times. Low density material is quickly accelerated between $t=5-20$ \tcc in the \turb cloud, leaving only a few high density cloudlets at late times. We quantify this rapid mass loss in Figure~\ref{fig:mass_evolution}, which shows the normalized cloud mass as a function of cloud crushing time. The mass is calculated using material at or above 1/3 the initial median density. By $t=20$ \tccns, nearly 80\% of the original mass of the \turb cloud has been mixed into the hot wind and fallen below the density threshold of $n = 0.33$ $\mathrm{cm}^{-3}$.

The loss of material proceeds more rapidly for \turb clouds than for spheres, even with the cloud-crushing time calculated as a function of the median density. As can be seen in Figure~\ref{fig:swn1_evolution}, panel 3, after the initial pancake stage, the spherical cloud compresses into a single core, with a small surface area and high column density in the wind direction. This morphology is a result of the original shock moving in the wind direction combined with the compression from shocks on the sides of the cloud, which push material toward the center. This high density core presents a smaller surface area for ablation of material than the many small high density regions in the \turb cloud (compare the third panels of Figures~\ref{fig:cwn1_evolution} and \ref{fig:swn1_evolution}). 
The single bow shock at the front of the spherical cloud protects the core, while also resulting in a pressure gradient through the cloud. Cloud elongation in the direction of the wind gradually breaks up the original core into smaller fragments, which each have individual bow shocks and lose material. However, as demonstrated in Figure~\ref{fig:mass_evolution}, the overall process is much slower for spheres than for \turb clouds. Only $\sim$40\% of the original spherical cloud material has been lost by $t=20$ \tccns.

\subsection{Median Density and Cloud Lifetimes}

Figures~\ref{fig:cwn1_evolution} and \ref{fig:swn1_evolution} show the evolution of clouds with a median density of $\tilde{n} = 1$ $\mathrm{cm}^{-3}$. We have also run four high resolution simulations with lower median densities. Models T05 and S05, the \turb and spherical clouds with a median density of $\tilde{n} = 0.5$, show a mass and morphology evolution similar to their higher density counterparts. Figure~\ref{fig:mass_evolution} shows that the mass loss as a function of cloud crushing time is nearly identical for the $\tilde{n} = 0.5$ $\mathrm{cm}^{-3}$ and $\tilde{n} = 1$ $\mathrm{cm}^{-3}$ clouds out to 25 \tccns. However, as the original median density continues to decrease, the evolution begins to follow a qualitatively different path. The original shock that passes through the cloud does not compress the gas enough for it to reach densities where it can cool efficiently. The warm cloud gas quickly gets rarified and accelerated with the hot wind, causing the mass-loss of the clouds to proceed much more rapidly at lower original median densities. This difference is clearly visible in the $\tilde{n} = 0.1$ evolutionary tracks in Figure~\ref{fig:mass_evolution}.

\begin{figure}
\includegraphics[width=1.0\linewidth]{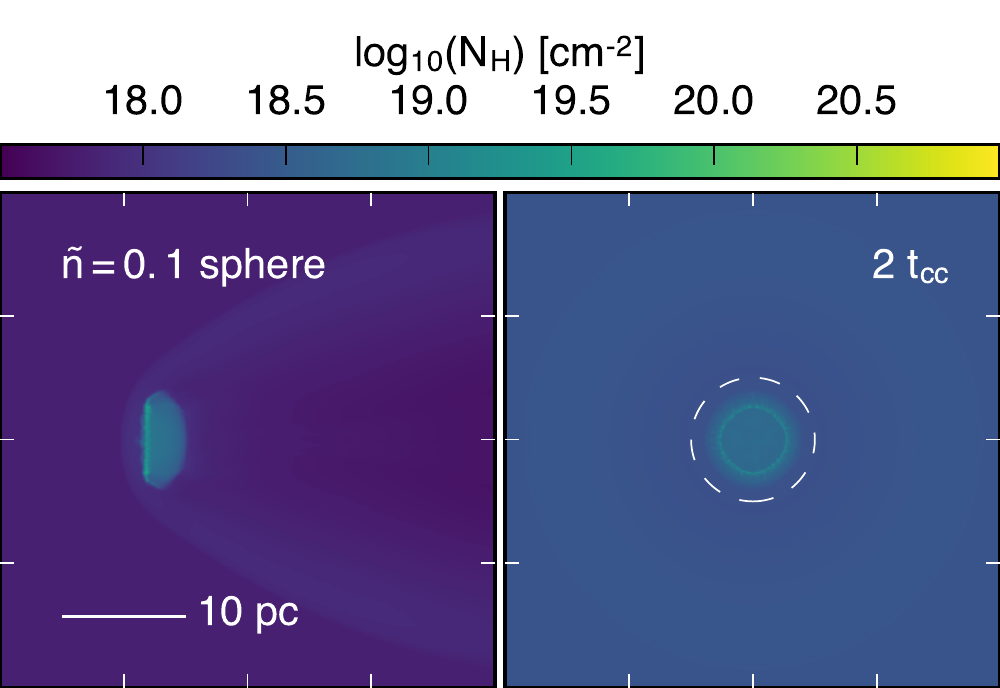}
\caption{Surface density projections of the $\tilde{n} = 0.1$ sphere simulation at $t=2$ \tccns. The left panel shows a $y$-projection and the right panel shows an $x$-projection, as in Figures~\ref{fig:cwn1_evolution} and \ref{fig:swn1_evolution}. A small ring of high surface density material is formed by the collision of shocks within the cloud, but most of the cloud remains at densities too low to cool effectively. The dashed circle in the right panel shows the original extent of the cloud.}
\label{fig:swn01_msd}
\end{figure}

In model S01, almost none of the gas reaches the requisite density to cool. In fact, only a small ring of gas within the cloud reaches the threshold density. This ring is visible in Figure~\ref{fig:swn01_msd}, which shows surface density projections of the $\tilde{n} = 0.1$ sphere simulation at 2 \tccns. The ring structure is caused by the shock moving in the wind direction colliding with the oblique shocks coming in from the sides of the cloud. While the densities in this collision do not get large enough to result in a single compact core as seen in the higher density models, this small amount of dense gas is enough to cause the extended tail seen in the mass evolution track of Figure~\ref{fig:mass_evolution}. However, most of the cloud is destroyed in less than 5 \tccns, consistent with adiabatic simulations of cloud-shock interactions \citep{Schneider15}.

The evolution of the $\tilde{n} = 0.1$ $\mathrm{cm}^{-3}$ \turb cloud (model T01) is less dramatically different as compared to its higher density counterparts. Although the median cloud density is below that required for effective cooling, the lognormal density distribution of initial cloud material spans a range up to $n \approx 10$ $\mathrm{cm}^{-3}$. As a result, parts of the cloud that were initially above the median density are still able to cool effectively after being shocked. As Figure~\ref{fig:mass_evolution} shows, the post-shock mass loss for the $\tilde{n} = 0.1$ \turb cloud proceeds faster than the $\tilde{n} = 1$ or $\tilde{n} = 0.5$ clouds, but more slowly than the $\tilde{n} = 0.1$ sphere. We expect that the speed of mass loss for \turb clouds would continue to increase as the initial median density is lowered, until eventually the evolution proceeded on an adiabatic track (shown by the dotted black line in Figure~\ref{fig:mass_evolution}).

\section{Phase Structure of the Wind}\label{sec:phase_structure}

In this section, we investigate the physical state of the gas in our simulations. The typical temperature of gas of a given density is useful in interpreting the cloud evolution. The density threshold for rapid cooling is also an important feature that determines whether any dense gas will survive for many cloud-crushing times. The high resolution of our simulations across the entire volume enables this study of the detailed phase structure of the gas as it evolves.

\subsection{Density and Temperature Structure}

\begin{figure}
\begin{centering}
\includegraphics[width=1.0\linewidth]{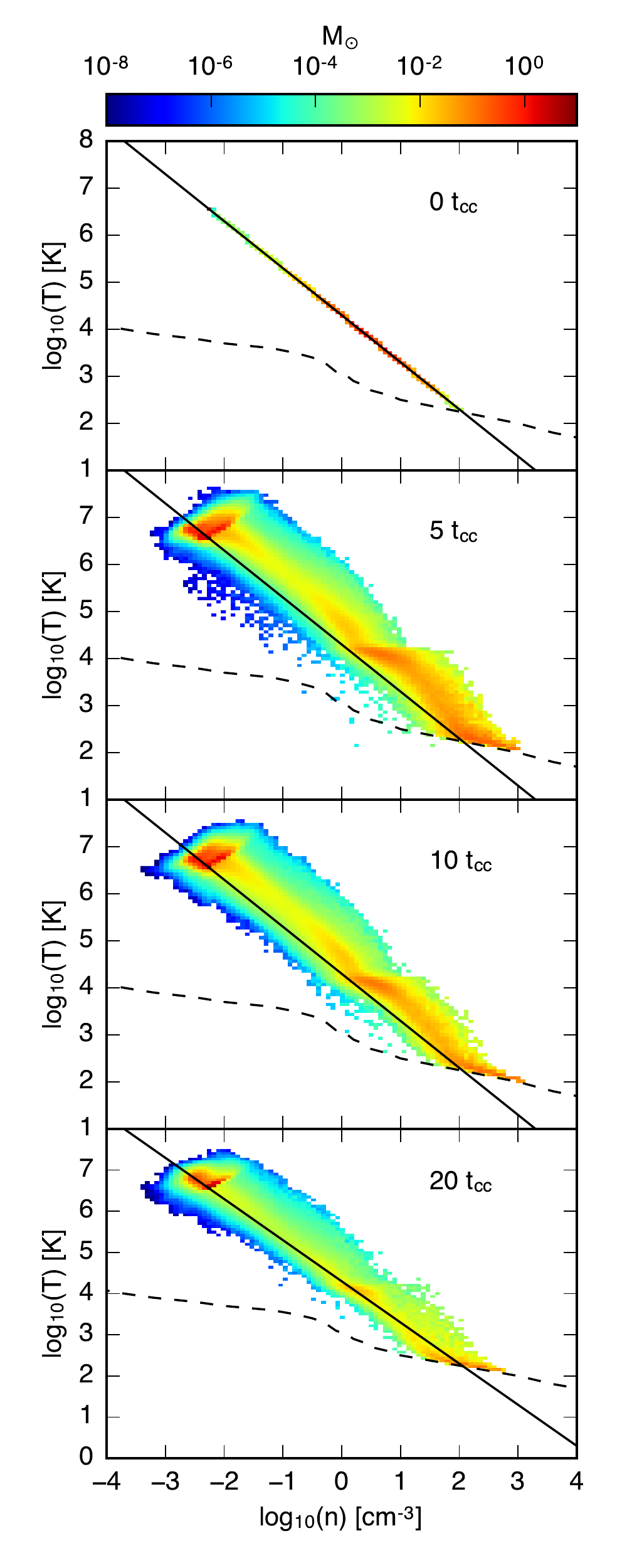}
\caption{Mass-weighted density-temperature phase diagrams at $t=0$, 5, 10, and 20 $t_\mathrm{cc}$ for the $\tilde{n} = 1$ cloud simulation (model T1). The initial equilibrium pressure is plotted as a solid line, and the temperature at which heating equals cooling as a function of density is plotted as a dotted line.}
\label{fig:cwn1_mnT}
\end{centering}
\end{figure}

Figure~\ref{fig:cwn1_mnT} shows density-temperature phase diagrams for the $\tilde{n} = 1$ \turb cloud at 4 different times in its evolution - the initial conditions, and $t=5$, 10, and 20 \tccns. Each bin is colored according to the total mass it contains, to better focus attention on the locations with the majority of the cloud material.
 Throughout the simulation, most of the mass is in the hot wind, visible as the red region at the upper left of the distribution in each panel (and as a single bin in the initial conditions). At $t = 0$ \tccns, the cloud's mass is distributed across a range of densities and temperatures, with most of the mass in regions at or above the median density of $\tilde{n} = 1$~$\mathrm{cm}^{-3}$. Each panel also contains an isobar showing the original pressure of the gas (recall that the cloud's pressure is matched to the wind in the initial conditions), as well as a dashed line that shows the equilibrium location between heating and cooling in our simulations.

After the cloud has been shocked, the density-temperature phase diagram takes on a characteristic shape. 
At early times, most of the gas is at higher pressure than the initial conditions. As the simulation proceeds, the cloud gas slowly evolves back toward 
thermal pressure equilibrium with the incoming wind. Cloud material quickly fills out the entire range of densities between the original cloud density and the wind. A large amount of mass has been compressed to high densities, much of it over an order of magnitude higher than the initial densities. The high density material cools effectively, with much of the cloud mass located just above the equilibrium cooling temperature, at the lower right in each panel. There also exists a mass concentration around $\mathrm{log}(n) = 0.5$ and $\sim2 \times 10^4$ K. This buildup reflects the shape of the cooling curve, with maximally efficient cooling around $10^5$ K and a steep falloff around $10^4$ K. These features remain throughout the subsequent evolution, as the mass in high density bins is gradually reduced.

\begin{figure}
\begin{centering}
\includegraphics[width=1.0\linewidth]{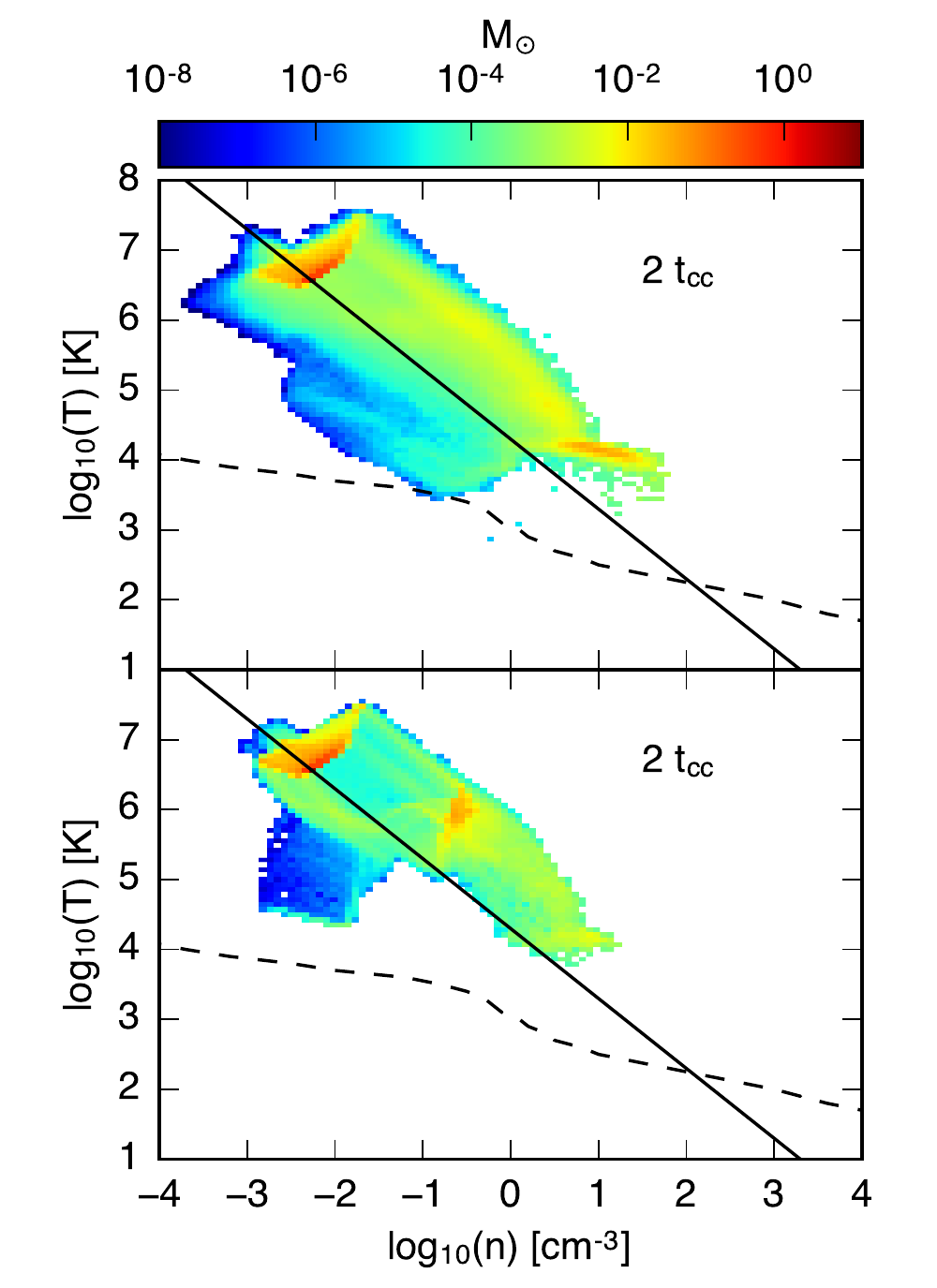}
\caption{Mass-weighted density-temperature phase diagrams for the $\tilde{n} = 0.1$ \turb (top) and spherical (bottom) cloud simulations at 2 \tccns. While much of the mass in the \turb cloud has crossed the $\mathrm{log}(n) = 0$ threshold and is able to cool efficiently, much of the mass of the spherical cloud remains below this density. The low-density gas is unable to cool and is quickly mixed into the hot wind.}
\label{fig:n01_mnT}
\end{centering}
\end{figure}

The phase diagrams for simulations T01 and S01 yield an estimate of the density threshold required for efficient cloud cooling in our simulations. Figure~\ref{fig:n01_mnT} shows the mass-weighted density-temperature phase diagrams for both simulations at $t=2$ \tccns. The \turb cloud, displayed in the left panel, has a large amount of mass in the cooling sweet spot just above $\mathrm{log}_{10}(n) = 0$. In contrast, most of the shocked sphere material never crosses this density threshold, and as a result, the gas remains at high temperatures and low densities where it quickly mixes with the hot wind. Animations of the phase diagrams show clearly the path that the cool gas takes in the \turb cloud simulation. Cloud gas originally above the median density shocks to densities just above $\mathrm{log}_{10}(n) = 0$, after which the gas begins to cool rapidly down to $10^4$ K. Only a small fraction of the sphere material reaches densities of $\mathrm{log}_{10}(n) = 0$, and as a result, most of the mass of the cloud is lost at early times.

\section{Momentum Coupling}\label{sec:momentum_coupling}

While the phase diagrams of the cloud gas are enlightening, in this section we seek to quantify several less obvious aspects of the cloud-wind interaction. The first area we address is cool cloud entrainment - the acceleration of cool material by the wind. We investigate entrainment in our simulations by studying 2D density-velocity histograms, to determine the typical velocities attained by gas of a given density. We follow our discussion of entrainment with an investigation of the detailed coupling of momentum between the hot wind and the cool cloud material. We assess how cloud mass transitions from one phase to another, and how the momentum in different phases changes over time. As in previous sections, we will focus on the fiducial $\tilde{n} = 1$ $\mathrm{cm}^{-3}$ \turb cloud simulation.

\subsection{Cool Cloud Entrainment}\label{sec:entrainment}

While cool clouds have been observed in galactic outflows at a variety of distances and velocities, the primary physical mechanism responsible for fast-moving cool gas continues to be debated. In this section, we investigate the velocity of the cool gas in our simulations, and show that in the purely hydrodynamic case, hot winds are unlikely to accelerate cool gas to the high velocities seen in many outflows.

\begin{figure}
\includegraphics[width=1.0\linewidth]{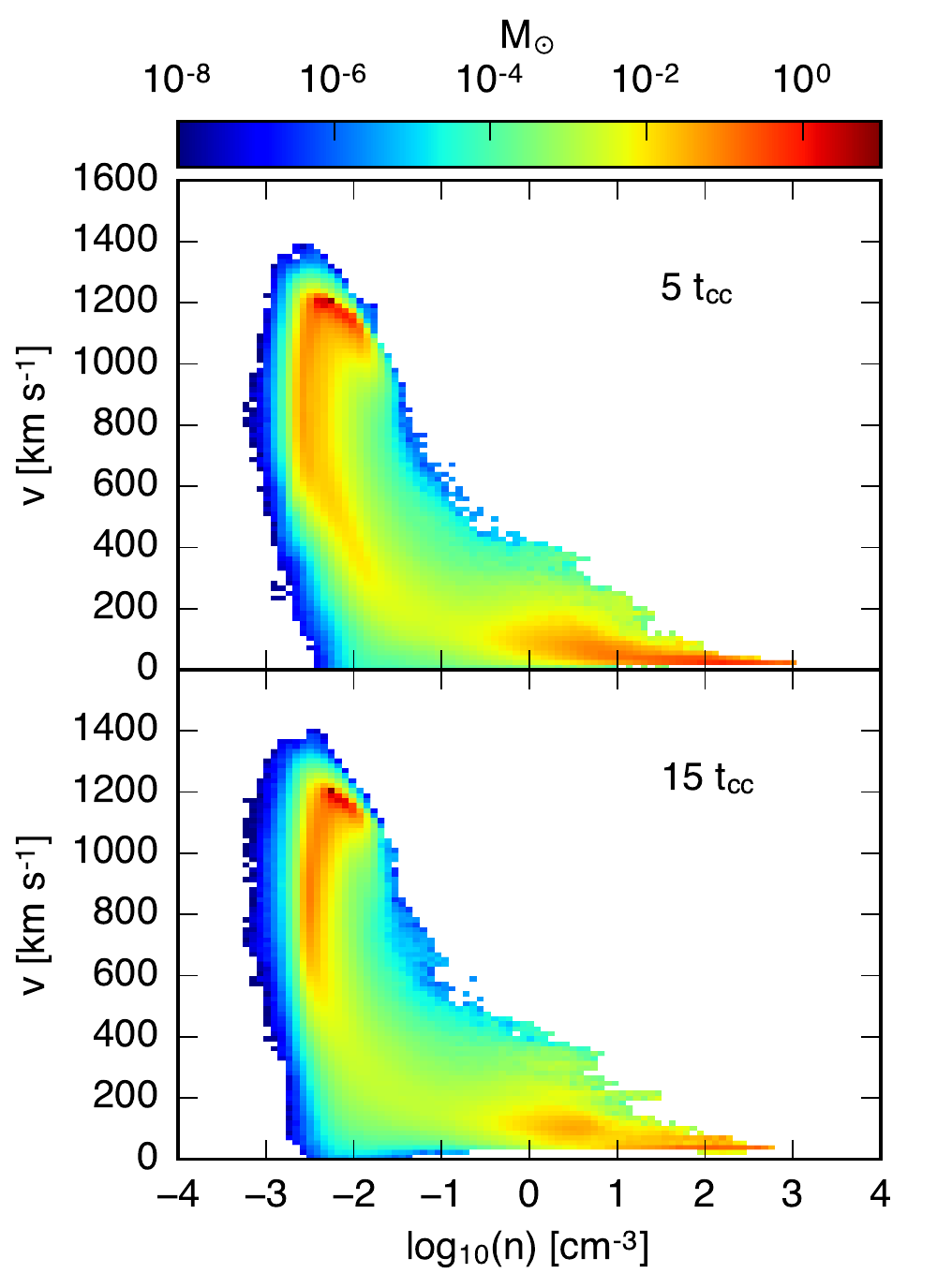}
\caption{Mass-weighted density-velocity phase diagrams at 5 and 15 $t_\mathrm{cc}$ for the $\tilde{n} = 1$ $\mathrm{cm}^{-3}$ \turb cloud simulation. High density material is not effectively accelerated in \turb clouds.}
\label{fig:cwn1_mnv}
\end{figure}

Figure~\ref{fig:cwn1_mnv} shows mass-weighted density-velocity histograms for model T1. The figure shows two representative snapshots, at $t=5$ and $t=15$ \tccns, with the velocity in the wind direction plotted on the $y$-axis. As in the density-temperature diagrams, the hot wind appears as a mass concentration at the upper left in each panel, with a small spread around the initial wind density and velocity. While cloud material has clearly acquired a range of densities, only low density material travels at high speeds. Only 3\% of the cloud mass above the original median density of $\tilde{n} = 1$ $\mathrm{cm}^{-3}$ is moving faster than 200 km $\mathrm{s}^{-1}$ by 15 \tccns. The average velocity of the dense material ($n > 1$ $\mathrm{cm}^{-3}$) is only $v_x \approx120$ km $\mathrm{s}^{-1}$.

\begin{figure}
\includegraphics[width=1.0\linewidth]{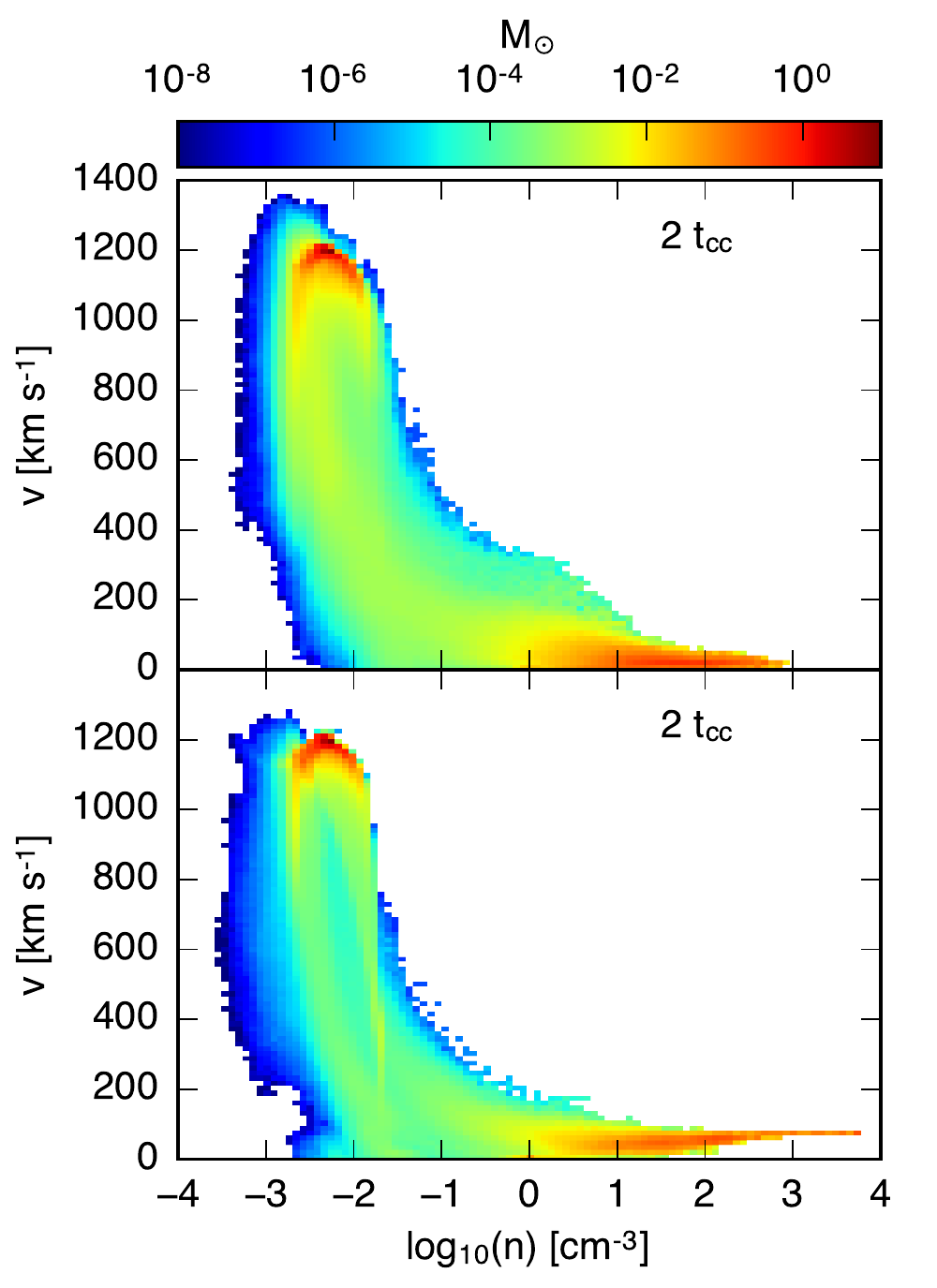}
\caption{Mass-weighted density-velocity phase diagrams for the $\tilde{n} = 1$ cloud (left) and sphere (right) at 2 $t_\mathrm{cc}$. Acceleration of dense gas is much more effective for spherical clouds, which has lower initial column density than the dense regions of \turb clouds with the same initial median density.}
\label{fig:cloud_sphere_mnv}
\end{figure}

One of the goals of this work was to investigate how different cloud density structures change the effectiveness of the hot wind in accelerating high density material. This question is addressed in Figure~\ref{fig:cloud_sphere_mnv}, which compares the density-velocity histograms for the $\tilde{n} = 1$~$\mathrm{cm}^{-3}$ \turb cloud at $t=10$ \tcc to the sphere. This figure indicates accelerating high density material proves more difficult in the \turb cloud case. This difference results from the different initial column densities of the cloud material, as we describe below.

In a simple model, the ram pressure of the hot wind will accelerate cold cloud material (we neglect other forces in the following analysis). The ram pressure of the hot wind scales as
\begin{equation}
P_\mathrm{ram} = n_\mathrm{wind} v_\mathrm{wind}^2,
\end{equation}
which gives a constant ram pressure per unit mass of $P_\mathrm{ram} / m_\mathrm{H} = 7.2 \times 10^{13}$~$\mathrm{cm}^{-1}\,\mathrm{s}^{-2}$ for our background wind conditions. The associated acceleration of the cloud, $g_\mathrm{cl}$, will then be
\begin{equation}
g_\mathrm{cl} = \frac{P_\mathrm{ram}}{m_\mathrm{H}}\frac{m_\mathrm{H}}{\Sigma_\mathrm{cl}} = \frac{P_\mathrm{ram}}{m_\mathrm{H}}\frac{1}{N_\mathrm{H}},
\end{equation}
where $\Sigma_\mathrm{cl}$ is the average surface density of the cloud, and $N_\mathrm{H} = n_\mathrm{cl} L$ is the column density of a particular region of the cloud. If the cloud is a constant density sphere, then $L$ can be approximated as twice the cloud radius. For the $\tilde{n} = 1$ $\mathrm{cm}^{-3}$ spherical cloud with a radius of $r\approx5$ pc, the column density is approximately $N_\mathrm{H} \approx 3\times10^{19}\,\mathrm{cm}^{-2}$ across the whole cloud, and the resulting acceleration is
\begin{equation}
g_\mathrm{cl} \sim 2.3 \times 10^{-6}\,\mathrm{cm}\,\mathrm{s}^{-2} \left[ \left(\frac{n_\mathrm{cl}}{1 \mathrm{cm}^{-3}}\right)\left(\frac{L}{10\,\mathrm{pc}}\right) \right]^{-1},
\end{equation}
or
\begin{equation}
g_\mathrm{cl} \sim 0.75\,\mathrm{km}\,\mathrm{s}^{-1}\,\mathrm{kyr}^{-1} \left[ \left(\frac{n_\mathrm{cl}}{1\,\mathrm{cm}^{-3}}\right)\left(\frac{L}{10\,\mathrm{pc}}\right) \right]^{-1}.
\label{eqn:cloud_acceleration}
\end{equation}
The cloud crushing time for the $\tilde{n} = 1$ clouds is \tcc$\approx56.4$ kyr. The acceleration in Equation~\ref{eqn:cloud_acceleration} gives a cloud velocity of $v_\mathrm{cl}~\sim~85\,\mathrm{km}\,\mathrm{s}^{-1}$ after 2 \tcc (113 kyr), which is roughly consistent with our results for the velocities of the densest gas after 2 \tccns, shown in the right panel of Figure~\ref{fig:cloud_sphere_mnv}. In fact, the average velocity of gas denser than $n = 500$ $\mathrm{cm}^{-3}$ at 2 \tcc for the $\tilde{n} = 1$ sphere is $77\,\mathrm{km}\,\mathrm{s}^{-1}$. After the cloud has been crushed Equation~\ref{eqn:cloud_acceleration} is no longer an adequate estimate of the cloud acceleration, because the column densities have increased by over an order of magnitude (compare the first and second panels in Figure~\ref{fig:swn1_evolution}).

In contrast, at $t = 0$ the densest sight lines through the $\tilde{n} = 1$ $\mathrm{cm}^{-3}$ \turb cloud have column densities as high as $N_\mathrm{H} = 2\times10^{20}\,\mathrm{cm}^{-2}$, almost an order of magnitude larger than the sphere (see the first panel in Figure~\ref{fig:cwn1_evolution}). At these column densities, the cloud acceleration is only $g_\mathrm{cl} \approx 0.11\,\mathrm{km}\,\mathrm{s}^{-2}\,\mathrm{kyr}^{-1}$, yielding a dense gas velocity of $v_\mathrm{cl} \sim 12\,\mathrm{km}\,\mathrm{s}^{-1}$ at $t=2$ \tccns. As a result, the regions of high column density are accelerated less in the \turb cloud, and the high density gas at $t=2$ \tcc is traveling considerably slower than in the spherical cloud case. The average velocity of gas with $n > 500$~$\mathrm{cm}^{-3}$ is only $v\approx16\,\mathrm{km}\,\mathrm{s}^{-1}$ at $t=2$ \tccns.

In both the \turb cloud and sphere simulations, much less acceleration occurs at late times than predicted by Equation~\ref{eqn:cloud_acceleration} (see right panel, Figure~\ref{fig:cwn1_mnv}). As mentioned previously, this minimal acceleration can be explained by the increased column densities that result from the cloud crushing. In both the \turb and spherical clouds, compression towards the center caused by the initial shock results in column densities at $t=2$ \tcc that are an order of magnitude higher than the original values (compare the first and second panels in both Figure~\ref{fig:cwn1_evolution} and \ref{fig:swn1_evolution}). Thus, we conclude that entrainment of cool material to high velocities in a hot wind remains very inefficient under these circumstances, and the problem only compounds for clouds with more realistic shapes and density distributions.

These simulations also show that the average surface density of the cloud, $\Sigma_\mathrm{cl} = M_\mathrm{cl} / \pi R_\mathrm{cl}^2$, as given in Table~\ref{tab:simulations} fails to adequately predict how much dense gas will be accelerated. This fact can be demonstrated by comparing the dense gas acceleration for the $\tilde{n} = 0.5$~$\mathrm{cm}^{-3}$ \turb cloud (not shown) and the $\tilde{n} = 1$ $\mathrm{cm}^{-3}$ sphere. The $\tilde{n} = 0.5$ $\mathrm{cm}^{-3}$ \turb cloud originally has a lower average surface density than the $\tilde{n} = 1$ sphere - $\Sigma_\mathrm{cl} = 0.11$ and $\Sigma_\mathrm{cl} = 0.13$ $\mathrm{M}_\odot$ $\mathrm{pc}^{-2}$, respectively. However, the spatial coherence of the densest regions within the \turb cloud leads to individual column densities several times higher than the average column density of the sphere, around $N_\mathrm{H} \approx 10^{20}\,\mathrm{cm}^{-2}$ as compared to $N_\mathrm{H} = 3\times10^{19}\,\mathrm{cm}^{-2}$. As a result, the average velocity of densest material in the $\tilde{n} = 0.5$ $\mathrm{cm}^{-3}$ \turb cloud simulation is $v_x \approx 30$ km $\mathrm{s}^{-1}$, less than half the speed of the dense gas in the $\tilde{n} = 1$ $\mathrm{cm}^{-3}$ spherical cloud simulation.

\subsection{Integrated Mass and Momentum}\label{sec:momentum}

\begin{figure*}
\includegraphics[width=0.5\linewidth]{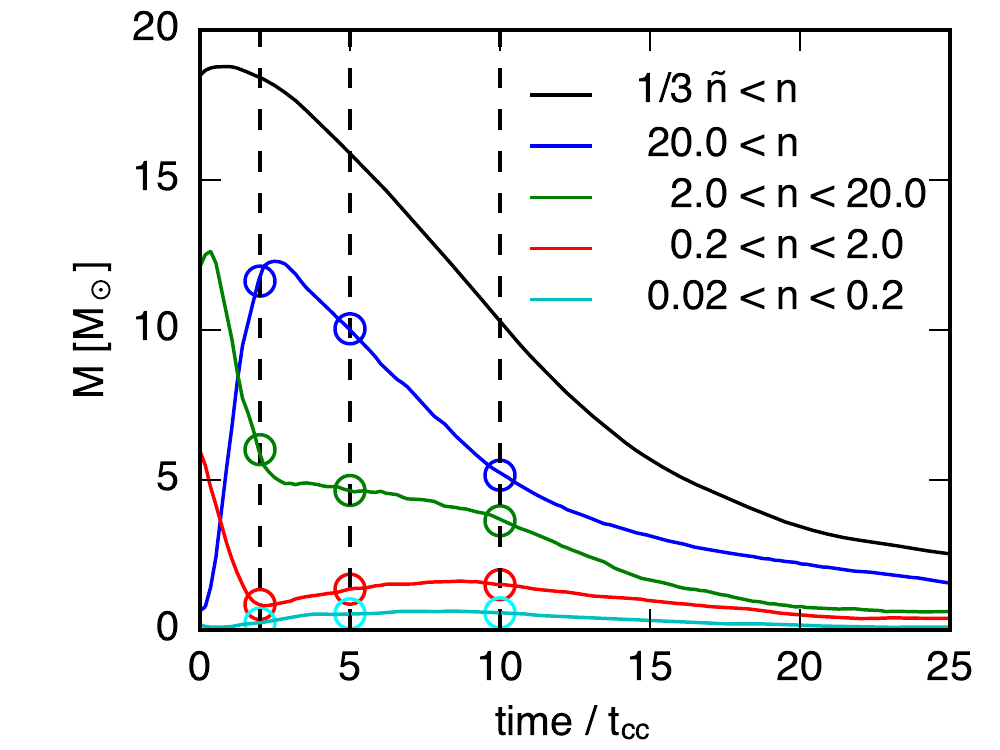}
\includegraphics[width=0.5\linewidth]{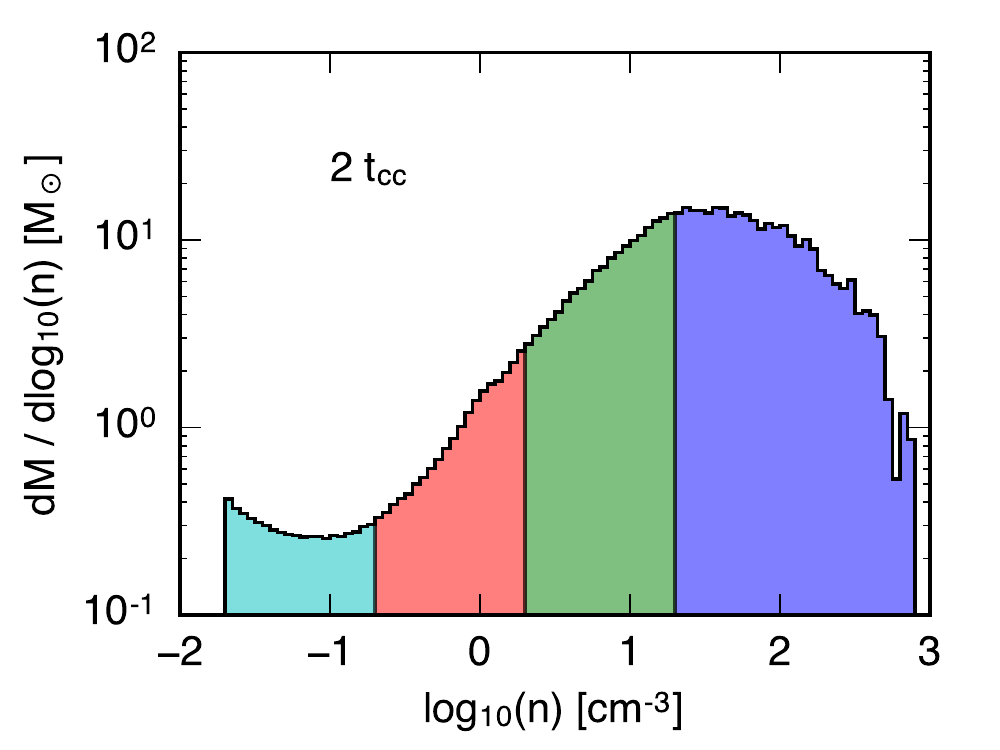}
\includegraphics[width=0.5\linewidth]{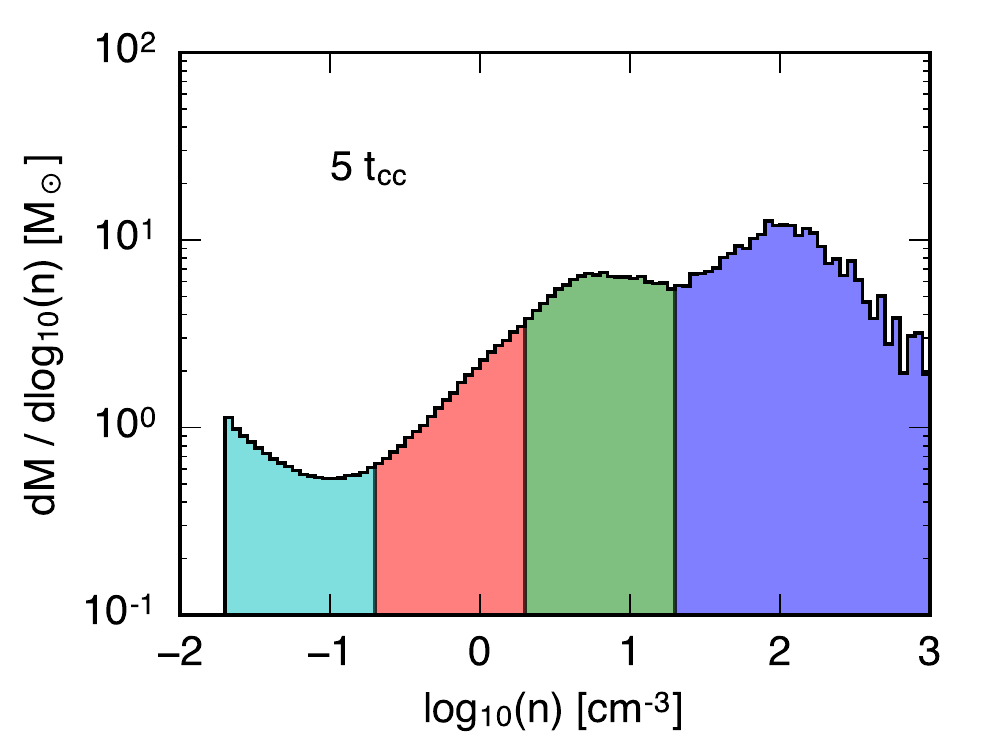}
\includegraphics[width=0.5\linewidth]{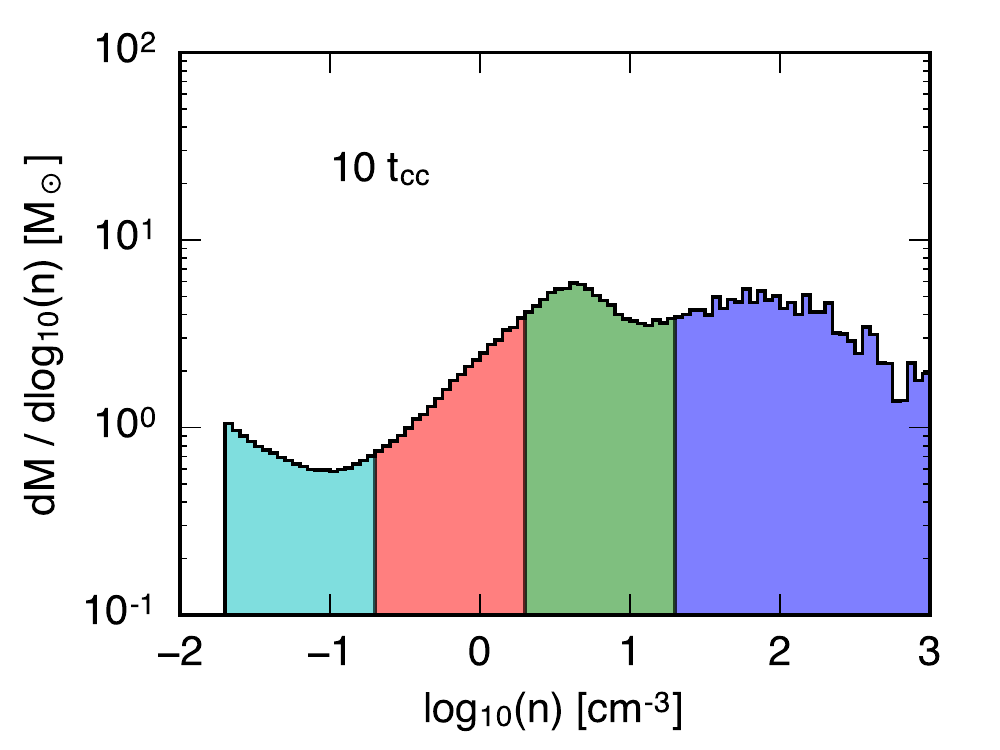}
\caption{Evolution of cloud mass in different density bins for the $\tilde{n} = 1$ $\mathrm{cm}^{-3}$ \turb cloud. In the upper left panel we show the total mass in different density bins; colors of the lines in the upper left panel correspond to density bins in the 1D histograms shown in the other three panels. Circles in the first panel correspond to the times displayed in the other three panels. The 1D histograms are normalized such that the integral over each density bin yields the total mass indicated by the circles.}
\label{fig:cwn1_mass_evolution}
\end{figure*}

Having established that in our simulations momentum does not transfer efficiently enough to accelerate the dense phase to the wind velocity, we would like to better quantify the momentum gained by other phases of the outflow. 
Figure~\ref{fig:mass_evolution} showed the evolution of cloud mass for each of our high resolution simulations as an integrated quantity, the total sum of material above a given density threshold. To better understand the way that gas evolves within the cloud, we now divide this evolution into multiple density bins. Figure~\ref{fig:cwn1_mass_evolution} shows this binned mass evolution plot for the $\tilde{n} = 1$ cloud. The black line matches the one displayed in Figure~\ref{fig:mass_evolution}, comprising all the material above a density threshold of 1/3 $\tilde{n}$ but no longer normalized by the initial cloud mass. Colored lines in Figure~\ref{fig:cwn1_mass_evolution} show the evolution of total cloud mass in four density bins: low, from $0.02 < n < 0.2$ $\mathrm{cm}^{-3}$; medium, from $0.2 < n < 2.0$ $\mathrm{cm}^{-3}$; high, from $2.0 < n < 20$ $\mathrm{cm}^{-3}$, and very high, $n > 20$ $\mathrm{cm}^{-3}$. The other three panels show 1D histograms of the mass as a function of density. Integrating any of the colored regions in the histogram yields the value represented as a circle on the line of the same color in the top left panel.

Much of the evolution in Figure~\ref{fig:cwn1_mass_evolution} takes place early on, with the most drastic shift occurring before 2 \tcc as the cloud is crushed. Initially, much of the cloud mass (12 $M_\odot$) is in the high density bin, $2.0 < n < 20$ $\mathrm{cm}^{-3}$, with a significant amount (5.9 $M_\odot$) also in the medium density bin that surrounds the initial median density of $n = 1$~$\mathrm{cm}^{-3}$. Only 0.64 $M_\odot$ is at $n > 20.0$ $\mathrm{cm}^{-3}$ initially. As the cloud is crushed, almost all of the material increases by about an order of magnitude in density, such that at 2 \tcc the medium density bin has only 0.85 $M_\odot$, the high density bin has 6.0 $M_\odot$, and 11.6 $M_\odot$ is in the very high density bin. These values are highlighted by circles at 2 \tcc in the first panel of Figure~\ref{fig:cwn1_mass_evolution}. Very little mass is in the low density bin at this time, and no cloud mass has been lost.

\begin{figure*}
\includegraphics[width=0.5\linewidth]{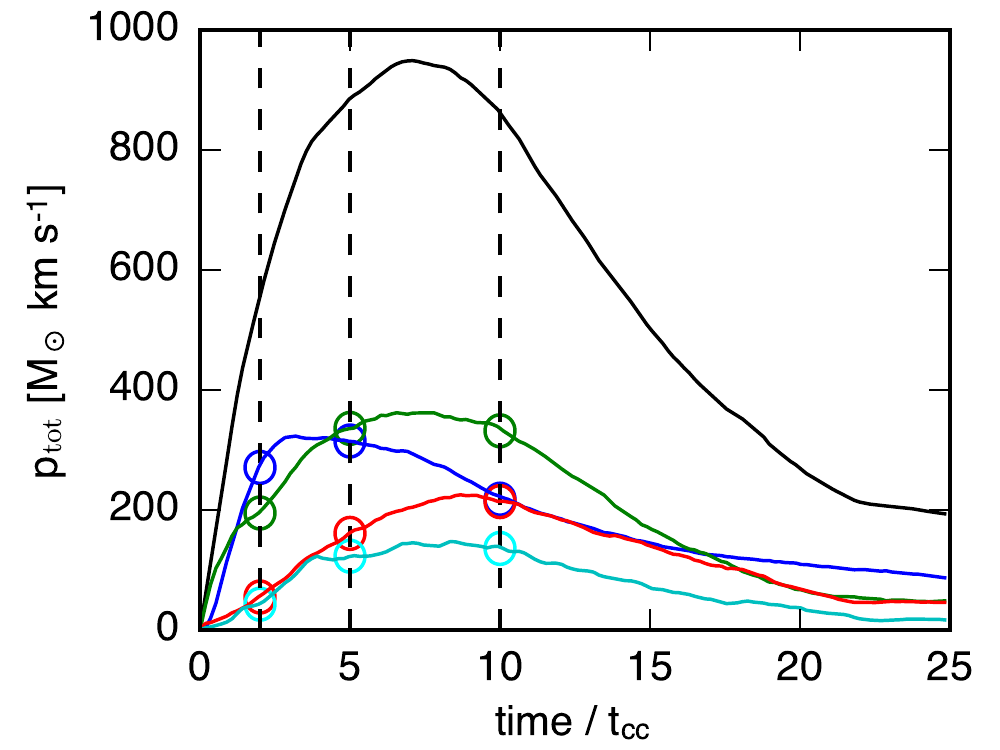}
\includegraphics[width=0.5\linewidth]{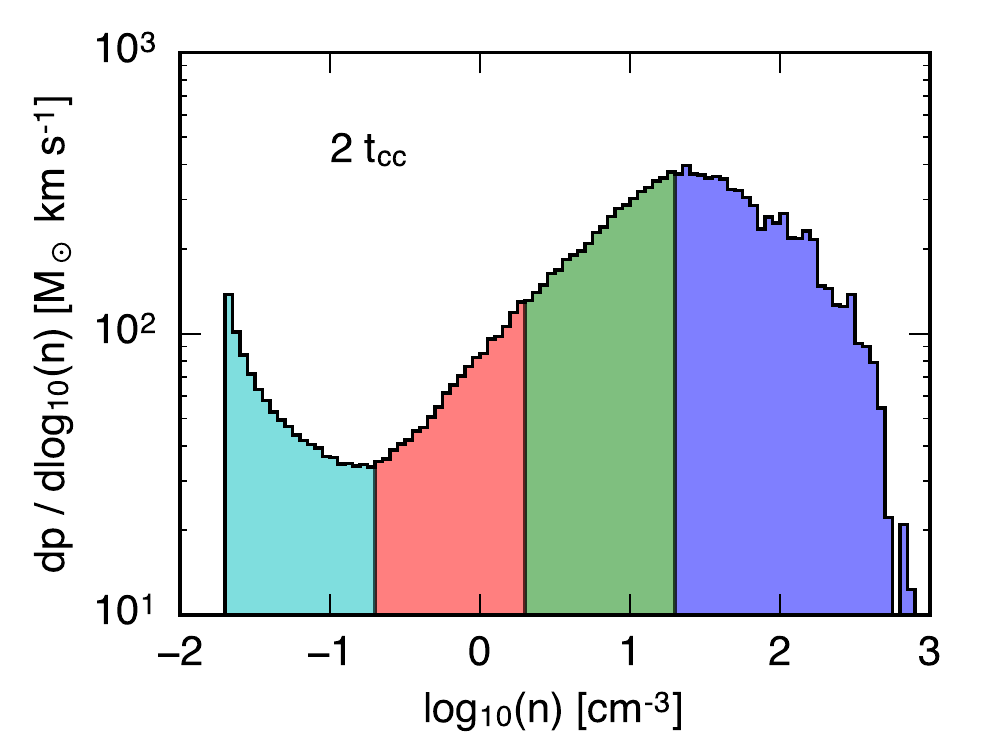}
\includegraphics[width=0.5\linewidth]{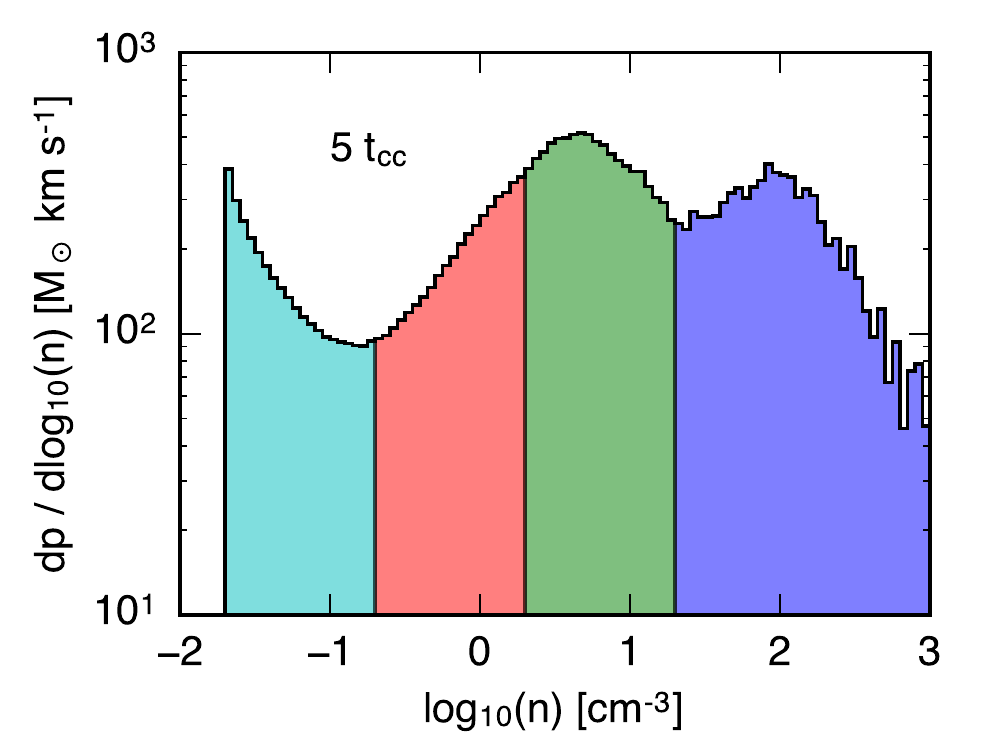}
\includegraphics[width=0.5\linewidth]{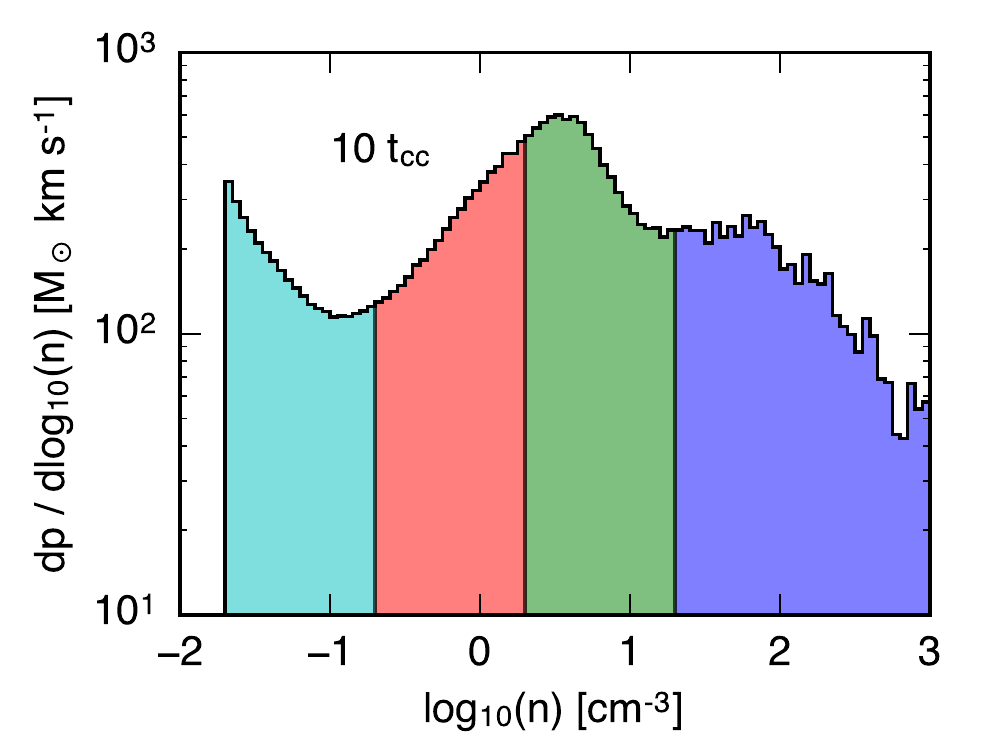}
\caption{Evolution of total momentum in the wind direction in different density bins for the $\tilde{n} = 1$ \turb cloud. The first panel shows the total momentum as a function of cloud crushing time, calculated by summing $M v_{x}$ over all cells with a density in the given range (see Eqn~\ref{eqn:momentum}). Colors represent the same density bins shown in Figure~\ref{fig:cwn1_mass_evolution}. Other panels show 1D histograms of momentum as a function of density at 2, 5, and 10 \tcc. Integrating under the histograms gives the values shown as circles in the first panel.}
\label{fig:cwn1_momentum_evolution}
\end{figure*}

After 2 \tccns, the evolution proceeds more gradually, with material slowly moving from higher density bins to lower ones. At 5 \tcc mass is concentrated in the same locations evident in the density-temperature phase diagram - there is a significant amount of mass (10.0 $M_\odot$) at very high densities, as well as a bump around $\mathrm{log}_{10}(n) = 0.5$ as a result of the shape of the cooling curve. These features are still present at 10 \tccns, though the mass in the highest density bin has been substantially reduced. At all times in our simulation after 2 \tccns, the majority of the cloud mass is in the highest density bin. Eventually we expect this mass to shift to lower density bins as the last of the high density gas is destroyed.

In the same way that we have integrated the total cloud mass, we can integrate total cloud momentum in the simulations. Figure~\ref{fig:cwn1_momentum_evolution} shows how the total momentum is divided amongst the same four density bins. The top left panel shows the evolution of the total momentum in the wind direction, computed as
\begin{equation}
p_\mathrm{tot} = \sum_{n = n_\mathrm{low}}^{n = n_\mathrm{high}} M_i v_{x, i}
\label{eqn:momentum}
\end{equation}
for each cell, $i$, in the simulation, where $M_i$ is the integrated mass in that cell and $v_{x}$ is the direction of the wind. The sum is taken over the same density ranges shown in Figure~\ref{fig:cwn1_mass_evolution}. In the other three panels, we show histograms of the momentum as a function of density at $t=2$, 5, and 10 \tccns. Integrals of the histograms are again shown as circles on the colored lines in the top left panel. At $t=2$ \tcc the momentum is distributed in a similar way to the mass, with most of the momentum in the two highest density bins. At $t=5$ \tccns, however, the distribution of total momentum has shifted. The three lower density bins have continued to gain momentum, but the highest density bin actually loses some, despite the fact that the highest density bin still contains the majority of the cloud mass. At this point, the total momentum in the medium and high density bins ($0.2\,\mathrm{cm}^{-3} < n < 20\,\mathrm{cm}^{-3}$) is $p_\mathrm{tot} = 495$ $M_\odot$ km $\mathrm{s}^{-1}$, 50\% more than is in the highest density bin ($n > 20\,\mathrm{cm}^{-3}$), $p_\mathrm{tot} = 315$ $M_\odot$ km $\mathrm{s}^{-1}$.

\begin{figure}
\includegraphics[width=1.0\linewidth]{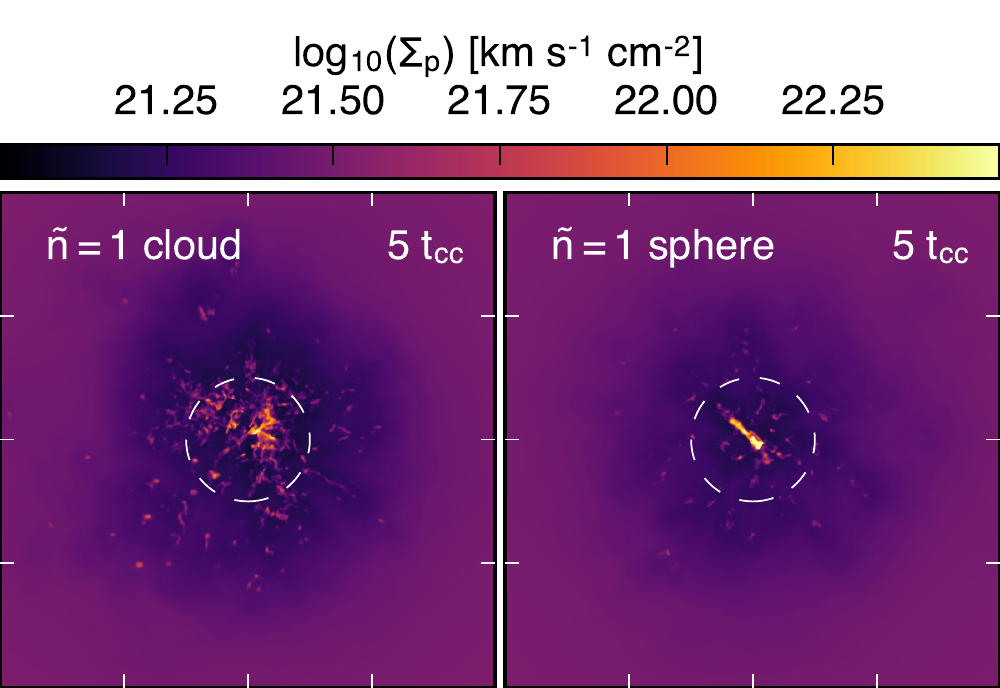}
\caption{Comparison of the momentum column density ($\Sigma_p = n v_x$ integrated in the wind direction) for the $\tilde{n} = 1$ \turb and spherical cloud simulations at $5$ $t_\mathrm{cc}$. The scale of the axes ticks is 10 pc. White dashed circles indicate the original extent of the cloud. Much higher momentum column densities are visible for the spherical cloud, which has been crushed to a single high density core.}
\label{fig:momentum_columns}
\end{figure}

This shift indicates that momentum is transferred efficiently from the hot wind to lower density gas. Gas with densities between $1\,\mathrm{cm}^{-3} < n < 10\,\mathrm{cm}^{-3}$ is in a relatively long-lived phase, giving it more time to gain momentum from the hot wind. The total momentum in the warm phase likely also increases as gas from higher density bins moves to lower density. Gas in the highest density bin with the most momentum may be the most likely to decrease in density. Without tracer particles, this shift is difficult to quantify in our simulations, but the highest density bin clearly loses mass at every time $t>2$ \tcc and at least some high density material with significant momentum shifts to the lower density bins. Overall, the distribution of total momentum across the different gas phases in the cloud is remarkably equal. 
The density-velocity histograms shown in Figure~\ref{fig:cwn1_mnv} also indicate this equality, showing that lower density material moves more quickly than higher density material. At very late times, $t > 17$ \tcc the highest density bin again has the most total momentum - this reflects the lack of mass in the lower bins by this time.

We can also compare the distribution of momentum between the \turb and spherical clouds by looking at the integrated momentum in the wind direction, a ``momentum surface density":
\begin{equation}
\Sigma_p = \sum_{i = 0}^{i = N_x} n_i v_{x,i},
\end{equation}
where $N_x$ is the total number of cells in the wind direction, $n_i$ is the number density of the gas in cell $i$, and $v_{x, i}$ is the velocity in the wind direction in cell $i$. We show projections of this quantity for the $\tilde{n} = 1$ \turb cloud and sphere at $t = 2$ \tcc in Figure~\ref{fig:momentum_columns}. As one would expect given the differences in the density-velocity phase diagrams, the momentum surface density is much higher for the sphere, reaching a maximum of $2.4\times10^{23}$ km $\mathrm{s}^{-1}$ $\mathrm{cm}^{-2}$ in the highest column. The momentum surface density of the background wind is $3.1\times10^{21}$ km $\mathrm{s}^{-1}$ $\mathrm{cm}^{-2}$.  In contrast, the momentum has a larger spatial extent and is spread over a wider range of gas column densities in the \turb cloud, with the maximum column of $4.3\times10^{22}$ km $\mathrm{s}^{-1}$ $\mathrm{cm}^{-2}$. Thus, even though the average momentum surface density at this time is similar between the two simulations, the momentum is clearly much more concentrated in the high density gas in the spherical cloud simulation. This result is consistent with the total momentum being shared across every density bin in Figure~\ref{fig:cwn1_momentum_evolution}.

\section{Resolution Effects}\label{sec:resolution}

At our fiducial resolution of 64 cells/$R_\mathrm{cl}$, we expect the global properties of cloud evolution to be qualitatively correct \citep{Gregori00, Poludnenko02, Melioli05}. These properties include the morphologies seen in the cloud destruction in Section~\ref{sec:cloud_evolution}, the general features seen in the phase diagrams in Section~\ref{sec:phase_structure}, and the multiphase distribution of momentum demonstrated in Section~\ref{sec:momentum}. However, many authors have suggested that a higher resolution of at least 128 cells/$R_\mathrm{cl}$ is required for convergence of quantitative measurements of properties like the mass loss rate and destruction time \citep{Klein94, Nakamura06, Scannapieco15}. The effects of resolution may be especially important for \turb cloud simulations, where increasingly dense structures are captured as the resolution of the simulation is increased. Therefore, we have carried out a numerical study to test the dependence of our results on the resolution of our simulations. We emphasize that this is \textit{not} a convergence study in the classical sense, because the initial conditions of the cloud change with resolution (as described below).

\begin{figure*}
\includegraphics[width=1.0\linewidth]{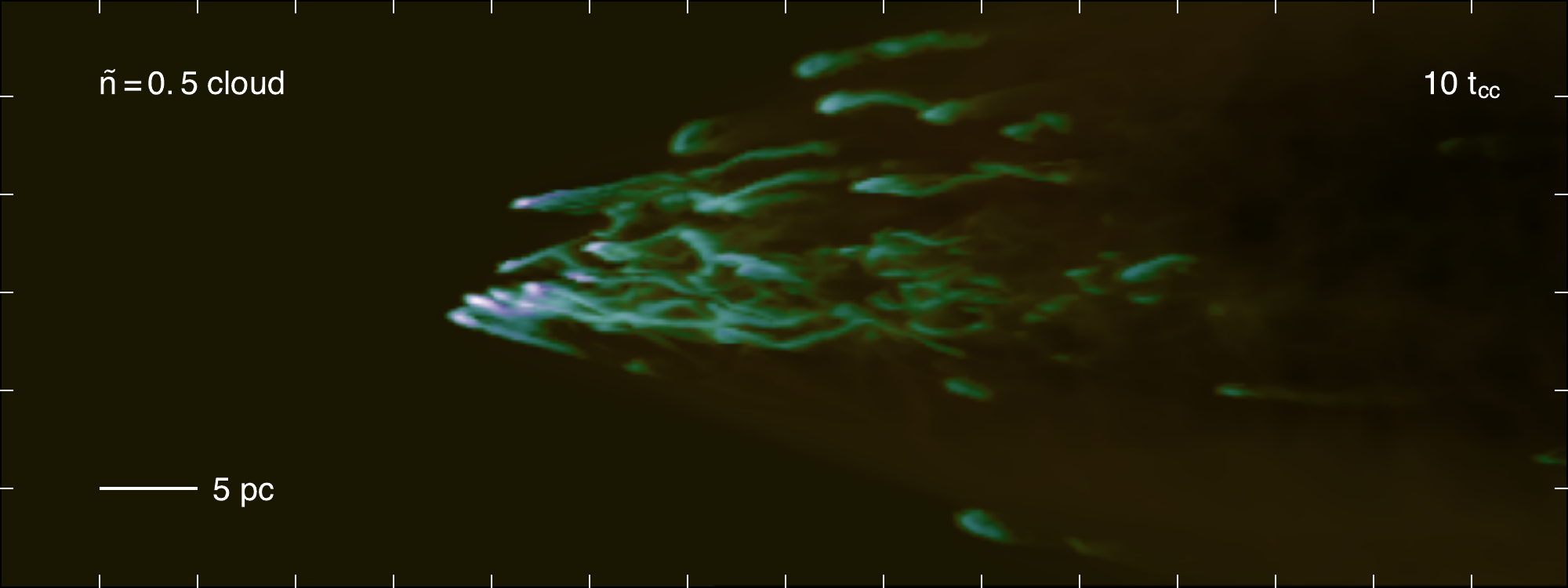}
\includegraphics[width=1.0\linewidth]{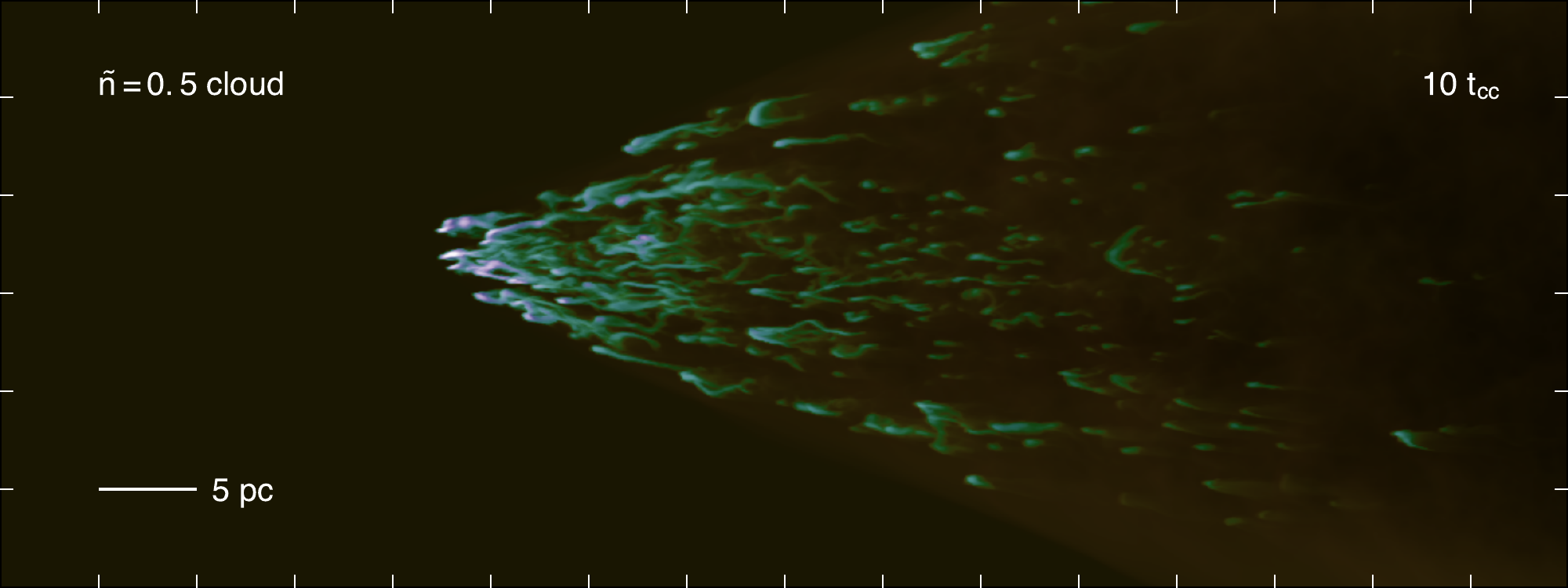}
\includegraphics[width=1.0\linewidth]{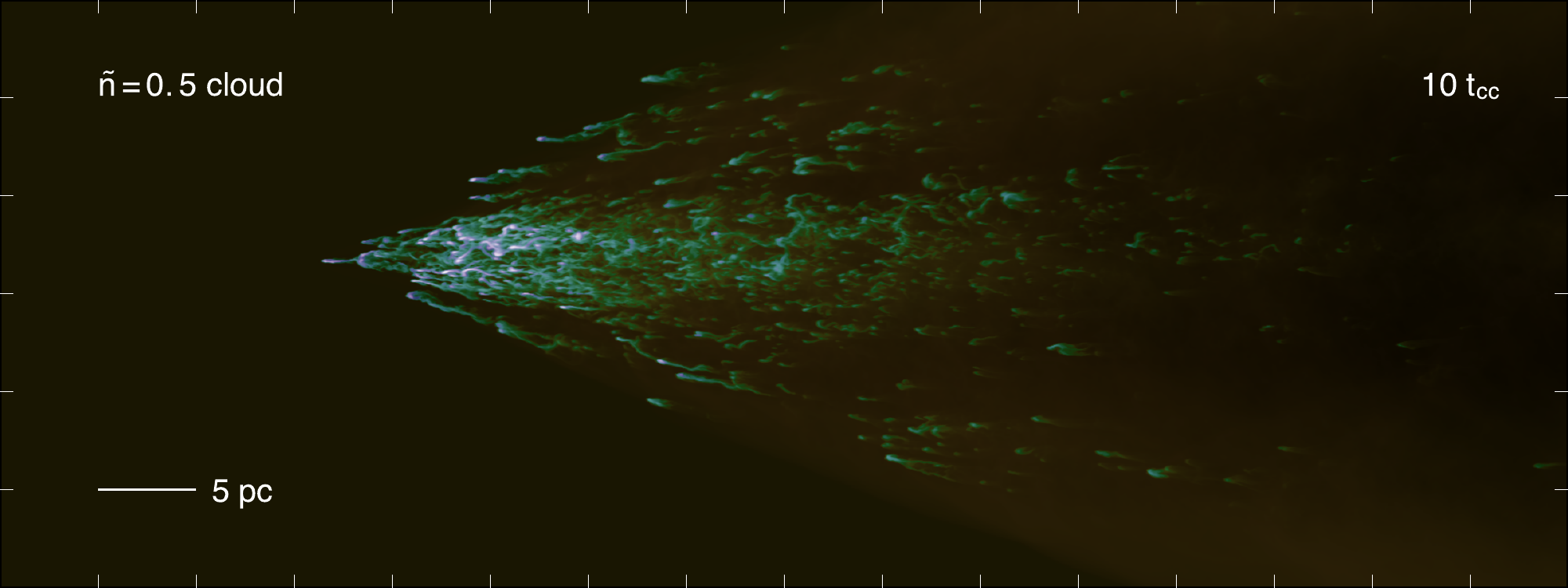}
\caption{A comparison at 10 \tcc between $\tilde{n} = 0.5$ \turb cloud simulations with three different resolutions: 32 cells/$R_\mathrm{cl}$, 64 cells/$R_\mathrm{cl}$, and 128 cells/$R_\mathrm{cl}$. The intensity in each panel corresponds to the projected number density, and color reflects the temperature of the gas (purple is cold, green is warm, and red is hot). As the resolution of the simulations is increased, the high-density features are resolved into smaller structures. As a result, the densest gas in the cloud is accelerated less efficiently with increasing resolution.}
\label{fig:resolution}
\end{figure*}

The study compares three versions of the $\tilde{n} = 0.5$ \turb cloud simulation: an ultra high resolution simulation run with 128 cells/$R_\mathrm{cl}$ (hereafter $R_{128}$), the production version with 64 cells/$R_\mathrm{cl}$ ($R_{64}$), and a low resolution version with 32 cells/$R_\mathrm{cl}$ ($R_{32}$). Because of the large computational expense of the $R_\mathrm{128}$ simulation, we use a physical box of size $80\times30\times30$ pc, as compared to the production simulations which ran in boxes of size $160\times40\times40$ pc. The $R_{128}$ simulation volume contains $2048\times784\times784$ cells, which yields a resolution of $\Delta x  = 0.039$ pc. The reduced time step required by the Courant condition and the computational expense of the additional cells mean that we can only afford to run the simulation for $\sim12$ \tccns. In practice, using the shorter box length results in the material leaving the computational domain at later times. The $R_{32}$ simulation is run in a box with the same physical size as the production simulation.

Each of the three simulations uses a cloud drawn from the same region of a Mach 5 isothermal turbulence simulation \citep{Robertson12}. To create the lower resolution clouds, the highest resolution simulation was resampled using a cubic spline interpolation. The bulk physical properties of the initial conditions are identical, including the median number density and total mass. As the resolution increases, we allow the cloud initial conditions to include progressively more small-scale structures and higher density regions. The initial conditions of the higher resolution simulations therefore do not simply reflect a better resolved version of low-resolution run initial conditions. The comparison presented below does not aim to act as convergence study, but instead attempts to capture how increasingly smaller scale features of the cloud initial conditions might influence the evolution of the wind-cloud system.

Figure~\ref{fig:resolution} shows snapshots of the three simulations at 10 \tccns, with the simulation time limited by the expense of the $R_{128}$ run. The intensity of the image in each panel scales logarithmically with the projected number density, and the color reflects the gas temperature. The $R_{32}$ and $R_{64}$ simulations have been cropped to display the same region as the $R_{128}$ simulation, for which the full box is shown. As expected, with each increase in resolution, finer-scale structure emerges. While in the $R_{32}$ simulation only $\sim 10$ cloudlets form, the increased resolution and the more detailed initial conditions of the $R_{128}$ simulation result in far more. At low resolutions the wind has successfully pushed the dense gas further, as evidenced by the bulk of the cloud material shifting further to the right in the upper panels. In the $R_{128}$ simulation, some of the dense gas has travelled less than $2 R_\mathrm{cl}$ in 10 \tccns.

\begin{figure}
\includegraphics[width=1.0\linewidth]{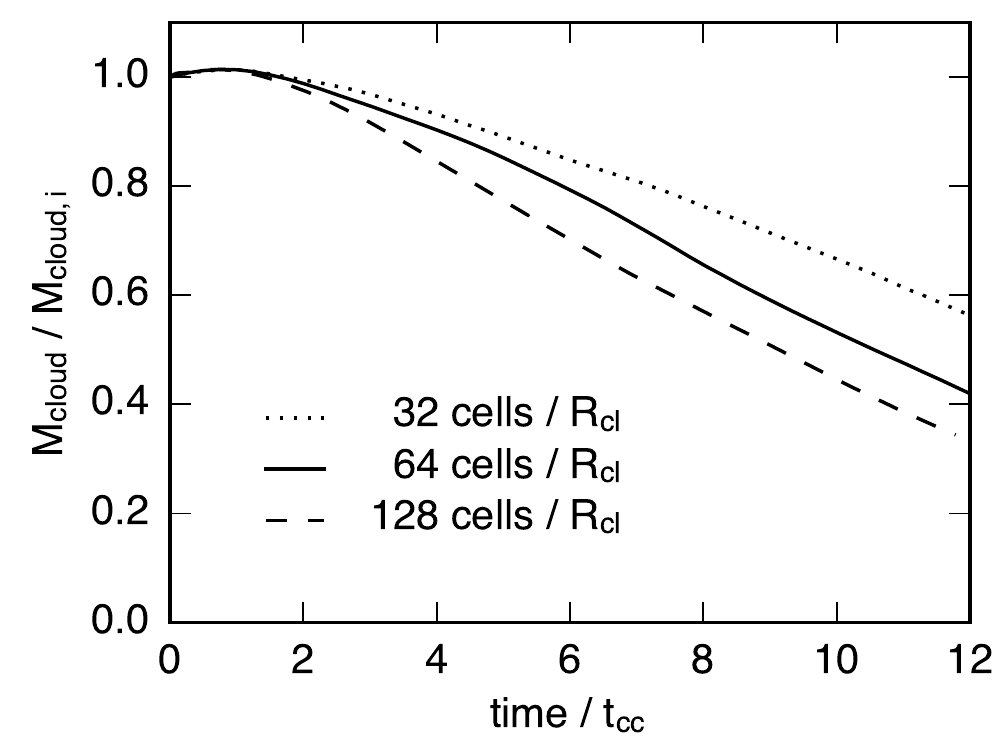}
\caption{Cloud mass is displayed as a function of time for the three $\tilde{n} = 0.5$ \turb clouds in the resolution study. The resolution of each simulation in terms of cells / cloud radius is displayed in the lower left. Mass loss initially proceeds more quickly for the higher resolution simulations.}
\label{fig:mass_convergence}
\end{figure}

\begin{figure}
\includegraphics[width=1.0\linewidth]{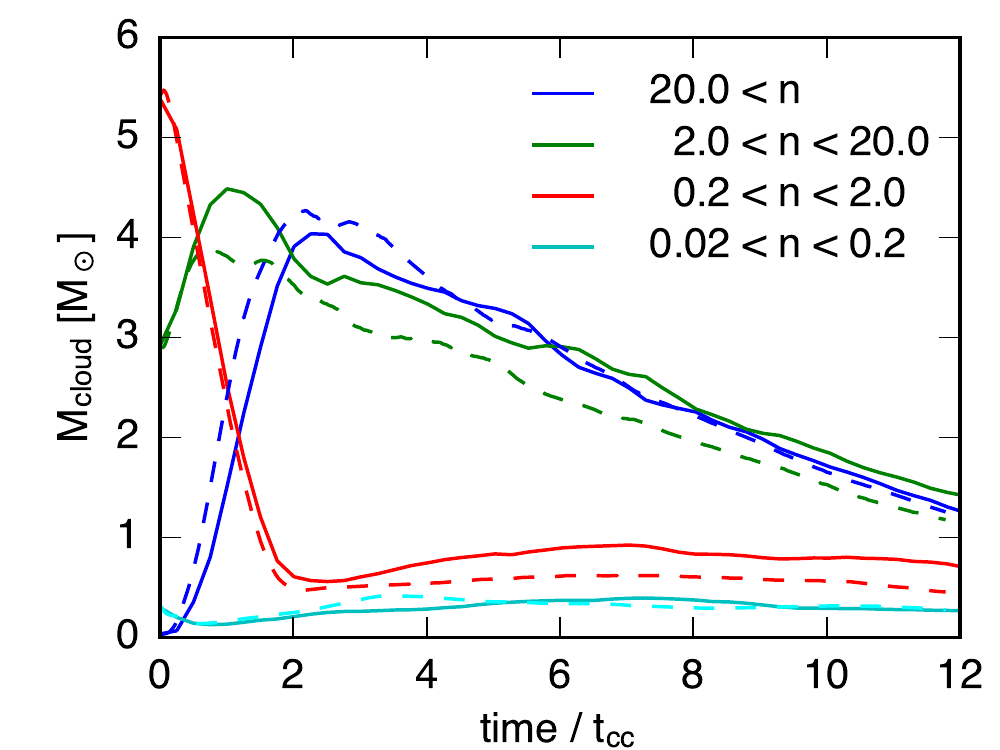}
\caption{The evolution of cloud mass in four density bins for the $R_{128}$ (dashed) and $R_{64}$ (solid) simulations. The decreased mass in the $2\,\mathrm{cm}^{-3} < n < 20\,\mathrm{cm}^{-3}$ density bin results in faster mass loss at early times for the higher resolution simulation, as the lower density cloud gas is quickly accelerated by the wind.}
\label{fig:cwn05_mass_evolution}
\end{figure}

For a more quantitative measure of the difference between these simulations, we plot in Figure~\ref{fig:mass_convergence} the mass evolution of each cloud. Even at our highest resolution, the results have not yet converged. Figure~\ref{fig:mass_convergence} shows a general trend toward more efficient mass loss as the resolution is increased. To better understand this trend, we examine the mass evolution in the separate density bins used in Section~\ref{sec:momentum}. The resulting mass-loss curves for the $R_{128}$ and $R_{64}$ simulations are shown in Figure~\ref{fig:cwn05_mass_evolution}. At early times, the primary difference between the two simulations appears in the two highest mass bins. The $R_{128}$ gains slightly more mass in the highest density bin ($n > 20\,\mathrm{cm}^{-3}$), but contains significantly less mass in the bin with $2\,\mathrm{cm}^{-3} < n < 20\,\mathrm{cm}^{-3}$ than the $R_{64}$ simulation.

At early times ($t < 2$ \tccns) the $R_{128}$ simulation generates less high density gas ($n > 1$ $\mathrm{cm}^{-3}$) - the density threshold required for efficient cooling. A higher fraction of the cloud gas in the $R_{128}$ simulation resides at densities and temperatures susceptible to quick destruction by the hot wind. This process of incorporation into the hot wind happens quickly, as evidenced by the lack of mass in the low density, slow ($v < 200$ km $\mathrm{s}^{-1}$) region in the density-velocity diagrams (see Figure~\ref{fig:cwn1_mnv}). Because cloud gas does not spend much time in this region of phase-space, the low density curves in Figure~\ref{fig:cwn05_mass_evolution} look similar. At later times, mass-loss proceeds similarly for the high resolution simulations (compare the slope of the blue and green curves after $\sim 5$ \tcc in Figure~\ref{fig:mass_convergence}).

\section{Discussion}\label{sec:discussion}

The results presented in the previous sections have important implications for the current theoretical understanding
of galactic winds. First, we have showed that the structure of the dense clouds plays a significant role in their evolution when exposed to a hot wind. Clouds that start with lognormal density distributions set by turbulent processes are mixed more quickly into the hot wind than spherical clouds with similar masses or median densities. This result implies that the mass-loading of galactic winds in the regions near the base of a hot outflow is an efficient process that likely results in significant mass-loading factors when hot gas interacts with denser clouds.

While our simulations indicate that mass-loading likely proves more efficient when the dense clouds have a turbulent structure, we have also demonstrated that this same structure in clouds tends to inhibit dense gas entrainment. The total acceleration of cloud material by the hot wind closely relates to the initial column density of the cool gas. In a \turb cloud, the spatial coherence of high density regions leads to large column densities along individual sight lines, which makes dense gas difficult to accelerate. While individual dense clumps can remain long-lived as a result of efficient cooling, dense gas ($n > 1$ $\mathrm{cm}^{-3}$) in our simulations rarely achieves velocities higher than $\approx 200$ km $\mathrm{s}^{-1}$. Cloudlets tend to get ablated and mixed into the hot wind before traveling more than $\sim30$ $R_\mathrm{cl}$.

These findings support a picture of galactic winds where mass-loading into the hot phase operates efficiently near the base of an outflow, and any surviving dense gas accelerates very little. However, this efficient mass-loading will result in hot winds that are more susceptible to thermal instability as they expand and cool. As calculated in \cite{Thompson16}, high mass loading factors will decrease the radius at which gas can cool out of the hot wind. If the gas that cools out of the hot phase retains significant velocity, it could explain the high velocity neutral gas seen at distances $1 - 100$ kpc from starbursting galaxies without a need to resort to entrainment of cool gas directly associated with the ISM. 

Furthermore, we have calculated in detail the distribution of mass and momentum associated with the clouds in our simulations. Far from being a simple two-phase medium, these winds are characterized by gas that spans a large range of densities and temperatures. Momentum from the hot wind couples to these phases with different efficiencies, such that the total momentum in each phase tends to be similar even if the mass is not. In cosmological simulations, where resolution limits the ability to capture these features of the multiphase galactic outflows, our calculations could be leveraged to improve treatments of the temperature and momentum distribution of the wind phases.

We now compare our findings to similar work in the literature. While many authors have studied the cloud-shock problem, relatively few have investigated the parameter space relevant to galactic winds and we will focus our attention on these results. We also discuss the potential effects that additional physics could have on our results, including different cooling rates, conduction, and magnetic fields. Finally, we include an analysis of the fate of the dense gas in our simulations in the presence of a gravitational potential.

\subsection{Cloud structure}

\cite{Cooper09} presented the only previous study that has investigated the destruction of clouds with a density distribution that is not symmetric along multiple axes. In that work, the authors compared the destruction of a radiatively-cooling fractal cloud to that of a radiatively cooling spherical cloud. The background hot wind properties in \cite{Cooper09} resembled ours. Our results agree well with theirs, in that they found that fractal clouds were more susceptible to fast destruction than spheres. However, their computational volume was too small to follow the evolution of the gas for many cloud-crushing times, and their resolution was relatively poor, and thus the fate of the small dense clumps that result from the destruction of an inhomogeneous cloud remained unclear. In our work, we find that while these small cores can survive for tens of cloud-crushing times as a result of efficient cooling, they are very difficult to accelerate due to their high column density. Additionally, higher resolution tends to amplify these effects. As a result, the cloudlets in our simulations are gradually destroyed over the course of $t\sim 20$ \tcc as the dense gas gets eaten away by the hot wind.

\subsection{Entrainment and Mass Loading}
\label{sec:discussion_entrainment}

In the past several years, several studies have investigated the ability of hot winds to entrain cool gas and carry it large distances from a galaxy. \cite{Scannapieco15} studied cloud-wind interactions using adaptive mesh refinement simulations with a maximum resolution equivalent to the fixed resolution of our simulations. The primary purpose of their work was to explore the effects that different background wind parameters had on cool cloud lifetimes and acceleration. Using spherical clouds of $T\approx 10^4$ K the authors derived a scaling for cloud destruction time that only depended on the mach number of the hot wind, with different density contrasts accounted for in the cloud-crushing time. Here, we compare our results to that scaling at $t_{50}$, defined as the time when 50\% of the cloud material is at or above 1/3 the initial cloud density. \cite{Scannapieco15} find
\begin{equation}
t_{50} = 4 t_\mathrm{cc} \sqrt{1 + M_{wind}},
\end{equation}
\label{eqn:mach_scaling}
which for our background wind would correspond to $t_{50} = 10$ \tccns. Looking at Figure~\ref{fig:mass_evolution}, we see that this actually fits the \turb clouds quite well, but our spherical clouds have a much longer lifetime. 

At first glance, this result may seem contradictory. However, the longer spherical cloud lifetimes may be explained by the different treatment of cooling in our simulations. As explained in Appendix~\ref{app:cooling}, we allow gas in our simulations to cool to temperatures as low as $T=10$ K using cooling rates calculated for solar metallicity gas. 
We also include the effects of a photoionizing background. The resulting heating keeps low density gas ($n < 1$ $\mathrm{cm}^{-3}$) warm, but is much less effective at higher densities. In contrast, \cite{Scannapieco15} simulated larger-scale clouds with the assumption of complete ionization, and therefore only allowed cooling above $T=10^4$ K. The ability of gas to cool to temperatures well below $T=10^4$ K in our simulations results in greater cloud compression. The resulting smaller surface area decreases the efficiency with which material ablates and correspondingly increases the cloud lifetime.

We find further evidence for this explanation by performing simulations with a different mean molecular weight $\mu$.
All the simulations presented in this paper used $\mu = 1$ when converting from mass density to number density,
\begin{equation}
n = \rho / \mu m_\mathrm{p},
\end{equation}
where $m_\mathrm{p}$ is the mass of a proton. However, we also performed a low resolution \turb cloud simulation with $\mu = 0.6$, the value appropriate for fully ionized solar metallicity gas. In comparing the lifetime of the cloud in this simulation with the same cloud in the standard simulation we find that the $\mu = 0.6$ cloud is destroyed more slowly, though the effect is small. This slight difference can be explained by the higher number densities for a given mass density in the $\mu = 0.6$ simulation. These higher number densities lead to more efficient cooling, which in turn leads to higher average densities at early times. The resulting high density clumps of gas prove more difficult for the hot wind to destroy than in the equivalent $\mu = 1$ model. Hence, the $\mu = 0.6$ cloud lives longer.

\cite{Scannapieco15} also find higher final velocities for their clouds than we do, even when comparing only spherical clouds. For clouds with similar background wind conditions they find velocities $v\sim 200-300$ km $\mathrm{s}^{-1}$ at $t=15$ \tcc (see their Figure 7), while we never see velocities above 200 km $\mathrm{s}^{-1}$. Again, this difference is likely a result of lower temperature gas in our simulations. Because our clouds can compress to higher densities in the initial stages of the wind interaction, their column densities increase more and acceleration is more difficult. This effect is compounded in the \turb cloud case with higher initial column densities.

\subsection{Additional Physics}

Other studies of cool gas in the context of galactic winds have investigated the effects of additional physics in cloud-wind interaction simulations. \cite{Bruggen16} perform a similar study to \cite{Scannapieco15} but incorporate the effects of conduction. They find that conduction can result in complete evaporation of the cloud at early times in cases where the column density of cloud material is below $N \simeq1.5\times10^{18}$ $\mathrm{cm}^{-2}$. The clouds we simulate do not start with column densities this low, and we therefore do not expect conduction to result in their rapid, complete destruction. We note the initial conditions of our hot wind most resembles their Mach 6.4 run that shows the least amount of difference in cloud evolution between the conduction and non-conduction models.

\cite{Bruggen16} demonstrate that when clouds in their simulations do survive, conduction tends to decrease the cross-section of the cloud presented to the hot wind. The decreased cross-section provides less surface area for ram pressure to act on the dense gas, which in turn decreases the efficiency of cloud acceleration. This effect is qualitatively similar to the impact of inhomogeneous column densities for \turb clouds described in Section~\ref{sec:entrainment}, though the origin is completely different. We suggest that the inhomogeneous cloud structure  may work in concert with conduction to further increase the early destruction of low column density material, and decrease the efficiency with which high column material accelerates.

Magnetic fields may also play a role in cool cloud evolution. A number of cloud-shock studies investigated the effects of planar magnetic fields in spherical clouds, with inconclusive results regarding whether the presence of field lines increased or decreased the time until cloud destruction \citep{Gregori00, Fragile05, Shin08}. More recently, \cite{McCourt15} performed wind-cloud simulations testing the effects of tangled magnetic fields incorporated within a spherical cloud. They showed that the presence of the magnetic field drastically increased the lifetime of the resulting cloudlets and increased their acceleration, allowing them to reach the hot wind speed without destruction. 

The \citet{McCourt15} simulations help motivate a future extension of our high-resolution wind-cloud simulations with realistic initial conditions to include MHD. The simulations by \citet{McCourt15} use a resolution of 32 cells/$R_\mathrm{cl}$. \citet{Scannapieco15} demonstrated that spherical cloud simulations with resolution less than 64 cells per cloud radius tend to prolong cloud lifetimes. Our simulations of \turb clouds show a trend toward faster mass loss with increasing resolution that has not yet converged at 128 cells/$R_\mathrm{cl}$. In addition, we have shown that spherical clouds have longer lifetimes than those with more realistic internal structure. While the results of \citet{McCourt15} suggest that including a tangled magnetic field would likely prolong the lifetime of the dense components of the \turb clouds we simulate, the combined effects of resolution and density structure make a quantitative prediction difficult. Incorporating magnetic fields in wind simulations with \turb clouds is an avenue that we aim to pursue in future work.

In addition to potentially important additional physics, parameters such as the metallicity of the cloud gas and nature of the background wind in our simulations will affect the quantitative results we have presented. In this work we have focused exclusively on the effects of cloud structure and surface density. The primary effect of changing the metallicity of the gas would be to change the cooling rates. As noted in Section~\ref{sec:discussion_entrainment}, tests with a different value of $\mu$ indicate that more rapid cooling leads to longer cloud survival, and our comparison with the simulations of \citet{Scannapieco15} indicates that less efficient cooling reduces cloud survival time. However, this is not an order-of-magnitude effect, and we therefore do not expect a change in metallicity to drastically alter our results.

On the other hand, as Figure~\ref{fig:wind_model} shows, the state of the background wind in the \citet{Chevalier85} model changes rapidly with radius as the wind escapes the galaxy. Given our choice of a single background wind state, we do not regard our simulations as completely generic with respect to galactic winds, though we do expect the general result of more rapid destruction of \turb clouds to hold. \citet{Scannapieco15} demonstrated a scaling with background wind mach number that indicates clouds survive longer with increasing mach number (see their Equation 22). In the future, we would like to investigate the relationship between \turb cloud lifetime and the background wind parameters, sampling a variety of distances from the galaxy that would inform the initial conditions for clouds at each distance.

\subsection{Ram Pressure vs Gravity}

As a final note, we consider the final fate of the dense gas in our simulations. Our simulations do not include gravity, and if the simulations continued running indefinitely eventually all of the cool material would mix into the outflowing hot wind. However, we can estimate the effect of the host galaxy's gravitational potential. Using an analysis similar to that in Section~\ref{sec:entrainment}, we can compare the expected acceleration of dense cloud regions owing to the wind's ram pressure, $P_\mathrm{ram}$, to their expected deceleration owing to gravity.

We can estimate the ram pressure acceleration as a function of column density, $N_\mathrm{H} = n_\mathrm{cl} L$, via
\begin{equation*}
g_\mathrm{ram} \sim 2.3\,\mathrm{km}\,\mathrm{s}^{-1}\,\mathrm{kyr}^{-1} \left(\frac{N_\mathrm{H}}{10^{19}\,\mathrm{cm}^{-2}}\right)^{-1}.
\end{equation*}
(This expression is equivalent to Equation~\ref{eqn:cloud_acceleration}.) Similarly, we can estimate the gravitational acceleration as a function of column density. At 1 kpc, the gravitational acceleration of M82 is
\begin{align}
g_\mathrm{grav} &\sim 1.4\times10^{-7}\,\mathrm{cm}\,\mathrm{s}^{-2} \left(\frac{M_\mathrm{M82}}{10^{10} M_\odot}\right)\left(\frac{R}{1\,\mathrm{kpc}}\right)^{-2}, \notag \\
g_\mathrm{grav} &\sim 0.044\,\mathrm{km}\,\mathrm{s}^{-1}\,\mathrm{kyr}^{-1}.
\end{align}
Given these estimates, we would expect cloud regions with column densities greater than $N_\mathrm{H} \approx 5\times10^{20}$ $\mathrm{cm}^{-2}$ to begin to fall back toward the galaxy. None of the clouds in our simulations initially have column densities quite this high - the densest sight lines in the $\tilde{n} = 1$ $\mathrm{cm}^{-3}$ \turb cloud reach $N_\mathrm{H} \approx 3\times10^{20}$ $\mathrm{cm}^{-2}$. Nonetheless, the acceleration due to ram pressure and deceleration due to gravity for these column densities are very similar, so in the presence of gravity the densest gas in our \turb cloud simulations would likely be accelerated very little or possibly fall back toward the central galaxy.

\section{Summary and Conclusions}\label{sec:summary}

In this work, we have modeled the hydrodynamic evolution of radiatively cooling clouds in the context of galactic winds with very high numerical resolution. Our study investigated two main parameters relevant to cold cloud survival - the initial structure of the cool gas and the median density of the cloud. We varied the cloud structure in our simulations between a lognormal density distribution with large-scale structure as set by turbulent processes and an idealized spherical distribution of gas. The median densities of our clouds ranged from $\tilde{n} = 0.1 - 1.0$ $\mathrm{cm}^{-3}$. The median density affects the overall destruction time of the cool gas via the cloud crushing time, as well as the efficiency of cooling within the cloud.

We find that clouds with a \turb density structure are destroyed more quickly than clouds with a homogeneous spherical density distribution. This efficient destruction results in faster mass-loading of the hot wind, as intermediate- and low-density regions of \turb clouds are quickly heated, rarified, and accelerated to the hot wind velocity. The entrainment of dense gas within cool \turb clouds proves extremely inefficient, and much less efficient than for idealized spherical initial conditions. The varying column densities present in \turb clouds result in very little acceleration of the densest regions, which are the only regions that survive for many cloud-crushing times. These effects are amplified as the resolution of the simulations is increased and the clouds are allowed to become increasingly realistic. We therefore conclude that entrainment of \turb ISM clouds in hot supernova winds does not explain the neutral gas observed at large distances from starburst galaxies, unless other physical processes (such as magnetic fields) substantially alter the results from the hydrodynamic case.

We have also provided an extensive description of the phase structure of the gas in the wind. Shortly after being shocked the gas associated with the \turb clouds spreads over a large range of densities and temperatures, with the densest regions cooling down to temperatures of $T \sim 100$ K. Each phase of gas remains close to thermal pressure equilibrium with the hot ($\gg 10^6$ K) wind. Interestingly, though the majority of the mass remains in the densest phases ($n > 20$ $\mathrm{cm}^{-3}$) for much of the cloud evolution, the total momentum distributes fairly evenly across densities. Roughly the same amount of momentum transfers to cold neutral (100's of K), cool ionized ($\sim 10^4 K$), and warm ionized ($\sim 10^5$ K) gas.

\acknowledgments
We acknowledge inspiration from the 2014 and 2016 Simons Symposia on Galactic Superwinds: Beyond Phenomenology. Some simulations from this work were performed on the EL GATO system at University of Arizona, funded by a National Science Foundation grant No. 1228509 (PI Brant Robertson). This research also used resources of the Oak Ridge Leadership Computing Facility, which is a DOE Office of Science User Facility supported under Contract DE-AC05-00OR22725, via an award of computer time on the \textit{Titan} system provided by the INCITE program (AST 109 and AST 117). We thank the anonymous referee for a constructive report. We also acknowledge Romeel Dav{\'e}, J. Xavier Prochaska, Evan Scannapieco, and Todd Thompson for discussions and comments that improved this work.

\appendix

\section{Integration Method and HLLC Riemann Solver}\label{app:integrator_hllc}

In \cite{Schneider15}, we presented the \cholla implementation of the 6-step corner transport upwind (CTU) integration scheme originally described in \cite{Gardiner08}. The unsplit CTU integrator preserves symmetry and minimizes numerical diffusion, making the scheme a good choice for many magnetohydrodynamics problems (see \citealt{Stone08}). Despite performing well in the standard \cite{Liska03} tests, we found that the transverse flux correction step of CTU led to estimates for the 1D density and energy fluxes that produced negative densities and pressures after the final update, particularly when used to model multidimensional strong radiative shocks. In fact, our implementation of the CTU integration method always fails for the simulations in this work, regardless of the choice of interface reconstruction method or Riemann solver. Therefore, we instead employed a very simple, robust integration scheme for the simulations presented in this paper, described below.

\begin{figure}
\includegraphics[width=0.5\linewidth]{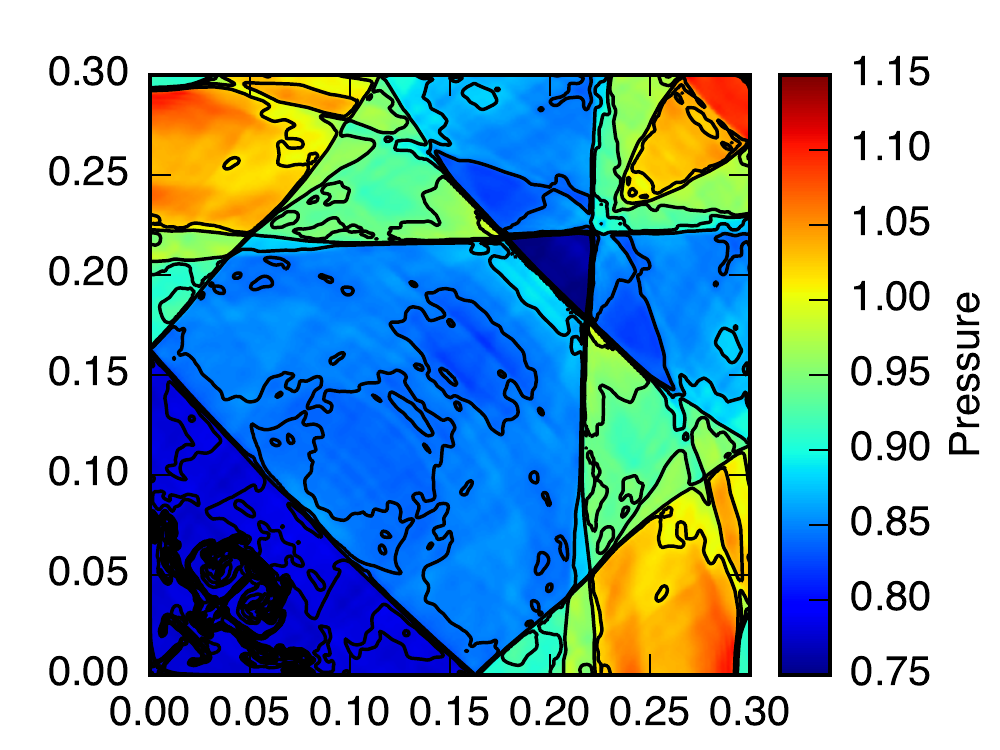}
\includegraphics[width=0.5\linewidth]{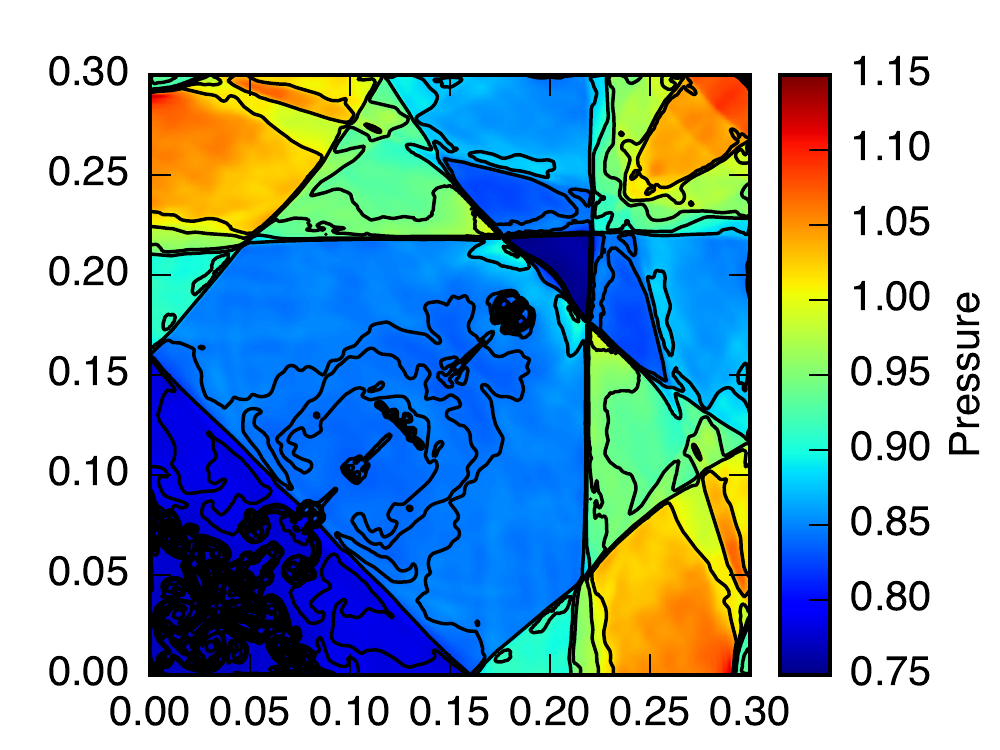}
\caption{A 2D implosion test, as presented in \cite{Schneider15}. A colored pressure map is overlaid with 36 density contours ranging from 0.125 to 1.0. On the left are the results at $t = 2.5$ for the integration scheme used for the simulations in the paper, on the right are the results at the same time using the CTU integrator. Both simulations used a CFL number of 0.1, piecewise parabolic reconstruction, and an HLLC Riemann solver. The additional diffusion in the integration scheme presented on the left prevents as clear a jet from forming, but also allows the \turb cloud simulations presented in this work to run without failing (which proved impossible with CTU).}
\label{fig:implosion}
\end{figure}

The integrator we use follows the initial steps of CTU, including piecewise parabolic reconstruction of the interface values in the primitive variables, characteristic time evolution of those interface values, and one-dimensional Riemann solutions at each interface to calculate fluxes. We implemented each of these initial steps as described in \cite{Schneider15}. However, rather than updating the interface values using the transverse fluxes, we simply skip to the final update and use only the one-dimensional fluxes to evolve the conserved quantities. Because the fluxes do not contain corrections for transverse directions, we found this method requires a very low CFL number to be stable - we use $cfl=0.1$ for all the simulations in this work. The low CFL number makes the integrator expensive despite its simplicity, but when combined with a dual energy formalism the method is quite robust. Figure~\ref{fig:implosion} shows the results of this new integration scheme as compared to the CTU method on the implosion test described in \cite{Liska03}. As the implosion test shows, this integrator is slightly more diffusive than CTU. The additional diffusion allows this simple integrator to succeed in circumstances where CTU fails. 

\begin{figure}
\includegraphics[width=0.5\linewidth]{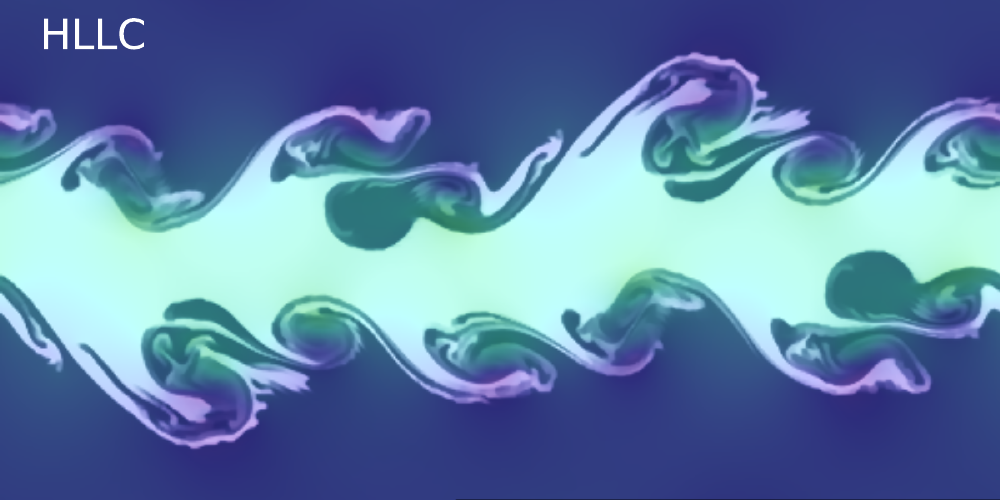}
\includegraphics[width=0.5\linewidth]{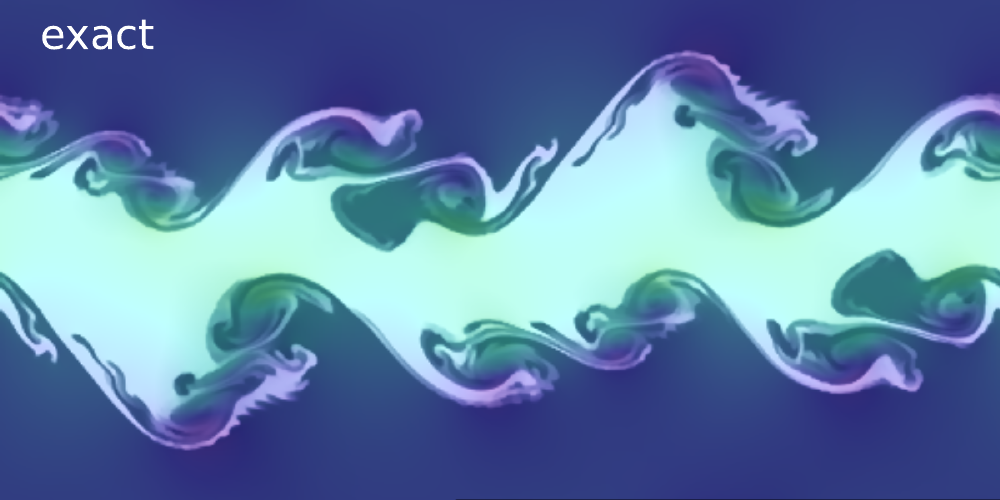}
\caption{We compare the results of a 2D Kelvin-Helmholtz simulation, starting with a discontinuous interface, on the left evolved using an HLLC Riemann solver, and on the right using an exact solver. Although the HLLC solver is an approximation, the contact discontinuity is well resolved with both methods.}
\label{fig:KH}
\end{figure}

We also employ an HLLC Riemann solver for this work \citep[see][]{Harten83,Toro94}. The \cholla implementation of the HLLC solver follows the description in \cite{Batten97}, which revises the original HLLC solver presented in \cite{Toro94} with updated maximum and minimum wave speeds that account for the possibility of colliding shocks within a cell. The HLLC solver has several advantages over the exact and Roe Riemann solvers presented in \cite{Schneider15}. Unlike the exact solver, the HLLC solver does not require an iterative procedure to calculate the intermediate-state pressure. This feature makes it a more natural fit to the GPU thread model, which works best on algorithms without thread divergence (i.e. fewer ``if" statements). Also, like all Riemann solvers in the HLL family, the HLLC is positive-definite \citep{Batten97}, meaning that it never produces negative pressures or densities in the intermediate state solution \citep{Einfeldt91}. These unphysical solutions can arise using linearized solvers like the Roe, which require ``fallback" techniques in failure scenarios \citep[see, e.g.,][]{Lemaster09}. Standard hydrodynamic tests show that the HLLC solver provides a comparable level of accuracy to the Roe and exact solvers, including problems that involve contact discontinuities \citep{Batten97}. We qualitatively illustrate this in Figure~\ref{fig:KH}, which compares well-resolved contact interfaces in Kelvin-Helmholtz simulations using the HLLC solver vs an exact solver.

\section{Dual Energy in \cholla}\label{app:dual_energy}

The conservative nature of Godunov methods enables shock capturing and allows them to correctly model shock heating, both highly desirable qualities in a hydrodynamics modeling scheme. However, in scenarios where the kinetic energy of the gas is large relative to the internal energy, the total energy formulation can lead to negative values of the internal energy. These scenarios may arise in the physical models we investigate here, where high mach number turbulent flows interact with rapidly cooling gas. A negative value of the internal energy does not necessarily affect the dynamic behavior of the gas in such scenarios, which is dominated by the much larger kinetic energy. However, an accurate estimate of the internal energy is required to correctly calculate the gas temperature, which is needed to determine the correct radiative cooling source terms. Therefore, in such scenarios, a modification of the total energy update may improve the physical realism of the model. So-called ``dual-energy" schemes that evolve both the internal energy and total energy simultaneously have proven to be a robust approach to fixing this problem \citep{Ryu93, Bryan95}. We employ a dual-energy formulation in \cholla that roughly follows the descriptions in \cite{Bryan14} and \cite{Teyssier15}. Below, we briefly outline the 1D version of the dual-energy update.

Without the dual-energy model, \cholla stores and evolves the conserved hydrodynamic variables: density $\rho$, the three components of the momentum vector, $\rho \mathbf{v} = (\rho v_x, \rho v_y, \rho v_z)^{\mathrm{T}}$, and the total energy, $E_\mathrm{tot} = \frac{1}{2} \rho \mathbf{v}^2 + e$, where $e$ is the internal energy. With the dual-energy model, \cholla also explicitly tracks an estimated internal energy. 
For the majority of steps in the hydrodynamic update, including interface reconstruction and Riemann solutions, the internal energy is treated as a passively advected scalar in the same manner as the transverse velocities. However, during the final update of the conserved variables \cholla evolves
the separately tracked internal energy in cell $i$ according to the following non-conservative equation:
\begin{equation}
e^{n + 1}_{i} = e^{n}_{i} + \frac{\delta t}{\delta x} \left[ (F(e)_{i - \frac{1}{2}} - F(e)_{i +\frac{1}{2}}) + \frac{1}{2} P_{i} (v_{i-1} - v_{i + 1}) \right].
\label{eqn:internal_energy_update}
\end{equation}
The second term on the right hand side of this update captures the flux of internal energy at the cell interfaces. The third term on the right hand side encompasses the change in internal energy due to pressure forces. We use $1/2 (v_{i-1} - v_{i+1}) / \delta x$ as a one-dimensional estimate of the velocity derivative. Both the pressure $P$ and velocities $v_{i-1}$ and $v_{i+1}$ used in Equation~\ref{eqn:internal_energy_update} are calculated at time $n$, making the update only first-order accurate in time. However, in cases where the conservative total energy update results in a negative estimate of the internal energy this first order estimate is more accurate.

Once the new total energy and internal energy have been calculated at time $n + 1$, we must determine which of the two internal energy calculations to use, the conservative estimate obtained by subtracting the kinetic energy from the total energy or the non-conservative estimate obtained with Equation~\ref{eqn:internal_energy_update}. In addition, the internal energy must be synchronized with the total energy after the update. Here we follow the decision tree outlined in  \cite{Bryan14}. At the end of the hydrodynamic update, the conservative estimate of the internal energy in cell $i$, $e_\mathrm{cons} = E_\mathrm{tot} - \frac{1}{2}\rho \mathbf{v}^2$, is compared to the the total energy in that cell. If the conservatively calculated internal energy is large enough (i.e., $e_\mathrm{cons} > 0.001 E_\mathrm{tot}$), we use the conservative estimate for the updated internal energy: $e^{n+1} = e_\mathrm{cons}$. In addition, to prevent the use of the nonconservative update in shocks, we compare the conservatively calculated internal energy to the maximum nearby total energy, $E_\mathrm{max} = \mathrm{max}(E_{i-1}, E_{i}, E_{i+1})$. If $e_\mathrm{cons} > 0.1 E_\mathrm{max}$, we again use the conservative estimate for the internal energy update,  $e^{n+1} = e_\mathrm{cons}$. However, if neither condition is met, we keep the non-conservative estimate for the separately tracked internal energy following Equation B1.

The last step in the dual-energy formulation is to synchronize the updated total energy with the updated internal energy. If the non-conservative estimate for $e^{n+1}$ is used, the value of the total energy must be corrected by subtracting off the old conservatively calculated energy, and adding the new non-conservative estimate, i.e. $E_\mathrm{tot}^{n+1} = E_\mathrm{tot}^{n+1} - e_\mathrm{cons} + e^{n+1}$.

\section{Optically-Thin Radiative Cooling in \cholla}\label{app:cooling}

The addition of radiative cooling requires the introduction of source terms to the right-hand side of the energy equation
\begin{equation}
\frac{\delta E}{\delta t} + \nabla \cdot [\mathbf{v} (E + P)] = \Gamma - \Lambda .
\end{equation}
Here, $E = \frac{1}{2} \rho \mathbf{v}^2 + e$ is again the total fluid energy per unit volume of the gas, $\rho$ is the density, $\bf{v}$ the velocity vector, and $P$ the gas pressure related to the internal energy $e$ via an ideal gas equation of state, $P  = (\gamma - 1) e$ with adiabatic index $\gamma$. The quantity $\Gamma$ is a source term that accounts for heating of the gas, and $\Lambda$ represents radiative losses \citep[see, e.g.,][]{Katz96}. In the following subsections, we describe how \cholla couples the source terms to the adiabatic equations and how we calculate $\Gamma$ and $\Lambda$.

\subsection{Coupling of Source Terms}
We implement radiative cooling in \cholla using an operator-split approach. After the hydrodynamic quantities have been updated for a given time step according to the ideal hydrodynamic equations, we add a source term to the internal energy to account for losses (or gains) as a result of radiative cooling (or heating) of the gas, such that
\begin{equation}
e^{n+1} = \tilde{e}^{n} + \dot{e} \Delta t_\mathrm{ad},
\end{equation}
where $\dot{e}$ is the rate of change of the internal energy, and $\Delta t_\mathrm{ad}$ is the hydrodynamic time step. Here, $\tilde{e}^n$ represents the updated internal energy after the hydrodynamic time step. The internal energy is calculated either from the gas pressure assuming an ideal gas equation of state, or tracked directly via the dual energy formalism. Because the adiabatic time step can be large compared to the radiative cooling time $e / \dot{e}$, the rate of change in internal energy is often a nonlinear function of the temperature over the course of a single $\Delta t_\mathrm{ad}$. Thus, we employ the common approach of subcycling the radiative cooling steps \citep[e.g.][]{Smith08, Gray10}, calculating a new $\dot{e}$ many times over the course of a single adiabatic time step so as to limit the change in internal energy for a radiative sub step, $\Delta t_\mathrm{rad}$, to less than five percent of the current internal energy:
\begin{equation}
\frac{\Delta{e}}{e} < 0.05.
\end{equation}
In practice, this update means that at the end of each hydrodynamic time step, we calculate the temperature for a given cell's number density and internal energy according to the ideal gas law
\begin{equation}
T = \frac{P}{n k},
\end{equation}
where $n$ is the number density of the gas in $\mathrm{cm}^{-3}$ calculated assuming $n = \rho / (\mu m_\mathrm{p})$, $m_\mathrm{p}$ is the mass of a proton, and $k$ is Boltzmann's constant. We have assumed a mean molecular weight of $\mu=1$ for all calculations, and have verified this does not influence our results
compared with values of $\mu=0.6$, as described in Section~\ref{sec:discussion}.

We next look up the tabulated net cooling rate associated with this density and temperature using a bilinear interpolation (described below), and calculate the resulting change in temperature over the full adiabatic time step given the cooling rate, $\Delta T$ = $\dot{T}$ $\Delta t_\mathrm{ad}$. If the change in temperature is greater than five percent, we shrink the radiative sub step such that $\Delta t_\mathrm{rad}$ results in a temperature change of five percent. We then update the temperature with this smaller time step, and repeat the process until we have synchronized with the full adiabatic time step.

\subsection{Calculating Cooling and Heating Rates}
\begin{figure}
\includegraphics[width=1.0\linewidth]{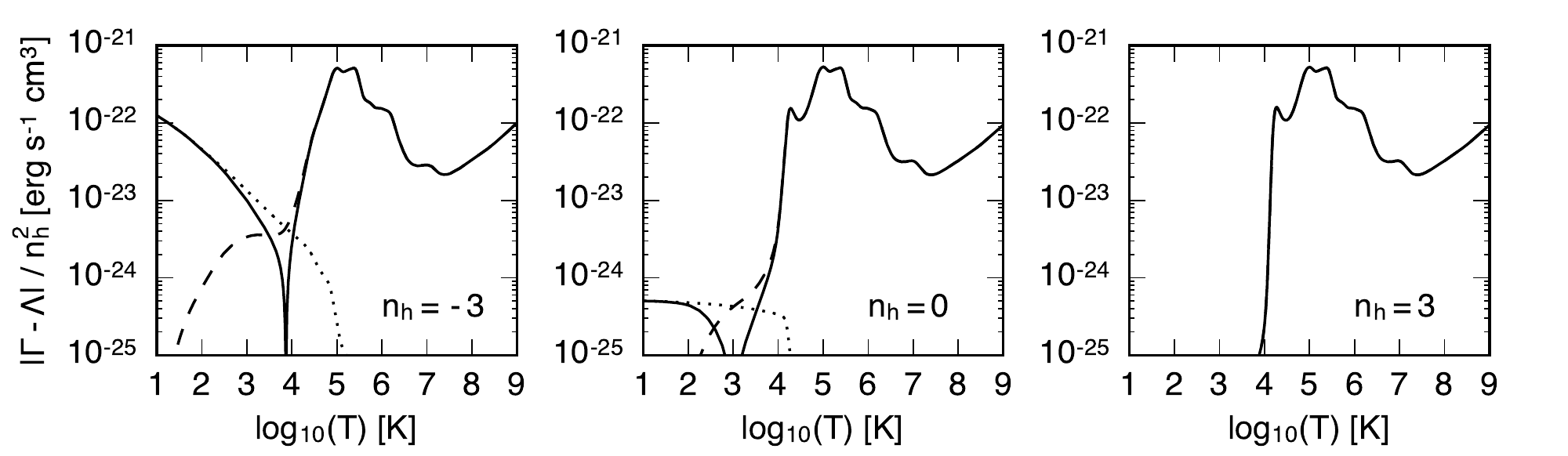}
\caption{Examples of cooling and heating rates as a function of temperature for solar metallicity gas calculated using Cloudy. The cooling function (dashed line), heating function (dotted line), and net energy gain or loss (solid) are shown for gas of three different densities. The gas is assumed to be in photoionization equilibrium while being exposed to the $z = 0$ CMB and a Haardt \& Madauu UV/X-ray background. At low densities, photoionization heating leads to an equilibrium temperature $\sim10^4$~K, while at higher densities heating is less effective.}
\label{fig:cooling_curves}
\end{figure}

We tabulate the cooling and heating rates per unit volume, $\Lambda$ and $\Gamma$ (measured in erg $\mathrm{s}^{-1}$ $\mathrm{cm}^{-3}$), using the Cloudy code, version 13.03 \citep{Ferland13}. For the calculations presented in this work, we assume that the gas is in photoionization equilibrium, subject to the $z = 0$ cosmic microwave background and the HM05 UV/X-ray background, as described in Hazy, the Cloudy documentation. The gas metallicity is assumed to be solar, using the GASS10 abundances in Cloudy \citep{Grevesse10}. Examples of the absolute cooling rates for gas of several different densities are shown in Figure \ref{fig:cooling_curves}.

Our cooling and heating rates are tabulated on a grid, with points calculated at equally spaced logarithmic intervals of  0.1 between 10 K $< T < 10^{9}$ K, and $10^{-6}$ $\mathrm{cm}^{-3} < n < 10^{6}$ $\mathrm{cm}^{-3}$. To perform the interpolation necessary to calculate $\Lambda$ and $\Gamma$ at an arbitrary $n$ and $T$, we take advantage of a novel feature of GPUs - texture memory. This special memory space allows a user to copy a 1, 2, or 3 dimensional array, or ``texture", onto the GPU, and then use built-in GPU functions to quickly retrieve arbitrary values from that texture using bilinear interpolation. We have tested this method against a similar CPU-based method using GSL bilinear interpolation functions, and find that the GPU texture memory approach typically speeds up the radiative cooling calculation by orders of magnitude as compared to a CPU function. In fact, when implemented using the operator-splitting approach described above for simulations run with $\sim128^3$ cells/GPU, the time spent calculating radiative cooling is completely negligible compared to the hydrodynamic calculations.

\bibliography{wind_paper}

\end{document}